\documentclass{article}

\usepackage{arxiv}

\usepackage[utf8]{inputenc} 
\usepackage[T1]{fontenc}    
\usepackage{hyperref}       
\usepackage{url}            
\usepackage{booktabs}       
\usepackage{amsfonts}       
\usepackage{nicefrac}       
\usepackage{microtype}      
\usepackage{graphicx}       
\usepackage{natbib}         
\usepackage{doi}            
\usepackage{xcolor}         
\usepackage{CJK}            

\usepackage{indentfirst} 

\hypersetup{
  colorlinks=true,
  linkcolor=red,
  citecolor=orange,
  urlcolor=blue,
}

\graphicspath{{./figure}}

\title{Three-dimensional decaying magnetic field \\ belonging to Beltrami flow
}

\author{\href{https://orcid.org/0000-0002-4875-8174}{\includegraphics[scale=0.06]{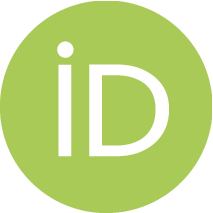}\hspace{1mm}Hideki Yanaoka (柳岡英樹)}
\thanks{Email address for correspondence: yanaoka@iwate-u.ac.jp} \\
	Department of Systems Innovation Engineering, \\
    Faculty of Science and Engineering, Iwate University, \\
    4-3-5 Ueda, Morioka, Iwate 020-8551, Japan \\
}

\renewcommand{\shorttitle}{Three-dimensional decaying magnetic field belonging to Beltrami flow}

\makeatletter
\hypersetup{
bookmarksnumbered,
pdftitle={\@title},
pdfsubject={},
pdfauthor={Hideki Yanaoka},
pdfkeywords={Navier-Stokes equations, Computational methods, Vortex dynamics, 
Magnetic fields, Turbulent transition},
}

\date{4 July 2022}
\makeatother

\begin{document}

\begin{CJK*}{UTF8}{ipxg} 
\maketitle
\end{CJK*}

\begin{abstract}
This study analysed a three-dimensional Taylor decaying vortex 
under an applied magnetic field as a benchmark test problem 
to verify the calculation method of an electromagnetic fluid flow 
and investigated the validity of the decaying magnetic field model. 
First, we observed the flow structure of a three-dimensional Taylor decaying vortex 
without an applied magnetic field. 
We investigated the changes in the error between the calculation result 
and the exact solution when the number of grid points and the Reynolds number varied 
and showed the effectiveness of the benchmark test. 
Next, we analysed a three-dimensional Taylor decaying vortex 
under an applied magnetic field and clarified the characteristics of the decaying magnetic field. 
When a magnetic field is applied, 
low magnetic pressure regions are connected in a mesh pattern, 
and the magnetic pressure distribution with a distorted cubic structure occurs 
to surround a high magnetic pressure region. 
In a stagnation region, the magnetic energy becomes low, 
and the magnetic flux line is similar to the streamline of the velocity field. 
High current densities occur in a grid pattern, 
and the magnetic flux lines swirl around the high current density region. 
The magnetic pressure and magnetic energy are high in the high current density region. 
When the Reynolds number and the magnetic Reynolds number vary, 
the decay trends of various energies agree well with the exact solution. 
The transition to turbulent flow occurs at a high Reynolds number, 
and the kinetic and total energies decrease rapidly. 
After the dissipation rate of kinetic energy becomes maximum, 
the vortex structure decays, and the flow field approaches a stationary state 
without magnetic fields. 
The three-dimensional Taylor decaying magnetic field belonging to the Beltrami flow 
is a valuable model for verifying the calculation method of electromagnetic fluid flows.
\end{abstract}


\section{Introduction}

\citet{Taylor_1923} derived an analytic solution for the two-dimensional flow field 
that satisfies the continuity equation and the Navier-Stokes equation. 
The solution represents a flow field in which the vortex decays, 
and the vortex is called a Taylor decaying vortex. 
This Taylor decaying vortex has been used to show the validity of a calculation method 
\citep{Kim&Moin_1985, Le&Moin_1991, Jordan&Ragab_1996}. 
An exact solution representing a three-dimensional Taylor decaying vortex 
has also been proposed \citep{Ethier&Steinman_1994, Barbato_et_al_2007, Antuono_2020}.

\citet{Ethier&Steinman_1994} derived an analytic solution 
for a three-dimensional decaying vortex 
using the method for deriving the solution for the two-dimensional Taylor decaying vortex. 
That solution is not periodic in the three directions of the coordinates, 
so it is necessary to model and give the boundary conditions. 
\citet{Barbato_et_al_2007} found periodic solutions in the three directions 
using the method of \citet{Ethier&Steinman_1994}. 
In that solution, each velocity component does not contain all the spatial coordinates 
of the independent variable, 
so the solution does not represent a complete three-dimensional flow. 
\citet{Antuono_2020} proposed an analytic solution of a periodic three-dimensional flow 
in which all the velocity components depend on the coordinates in the three directions, 
using the method of \citet{Ethier&Steinman_1994}.

Discretization of the fundamental equation is significant for the stability 
of numerical calculation in unsteady flow with a high Reynolds number. 
The conservative difference scheme allows for stable calculations 
and conserves kinetic energy in inviscid fluids. 
Since energy transformation occurs in the flow of electromagnetic fluid, 
it is not easy to verify the energy conservation characteristics 
in numerical analysis. 
Assuming zero kinematic viscosity and magnetic diffusivity, 
each volume integral of total energy and cross-helicity is conserved 
in a periodic flow \citep{Woltjer_1958}. 
Therefore, we can verify the conservation characteristics 
using these transport quantities for an inviscid periodic flow. 
It is hard to confirm whether energy is converted correctly in a viscous fluid. 
In addition, a difficult problem in an electromagnetic fluid analysis is 
to satisfy the constraint subjected to Gauss law for magnetism, 
that is, the condition of no divergence of magnetic flux density. 
When verifying a computational method for the flow of electromagnetic fluid, 
it is a suitable choice to use a model for which an analytic solution exists. 
Therefore, it is necessary to find an analytic solution 
for a three-dimensional flow under an applied magnetic field. 
If we can extend a mathematical model proposed by \citet{Antuono_2020} 
to a three-dimensional magnetic flow field, 
then we can use that model as a benchmark test. 
For a two-dimensional flow, \citet{Liu&Wang_2001} analysed the flow 
of the Taylor decaying vortex to which a magnetic field was applied.

In this study, we investigate a three-dimensional Taylor decaying vortex 
under an applied magnetic field and clarify whether that model is available 
for verifying the calculation method of an electromagnetic fluid flow.

\section{Fundamental equation}

The fundamental equations for the incompressible viscous flow of a Newtonian fluid 
under an applied magnetic field are the equations for mass, momentum, 
and magnetic flux density. 
The magnetic flux density must satisfy the constraint subjected to Gauss law 
for magnetism. 
These governing equations are given as
\begin{equation}
   \nabla \cdot {\bf u} = 0,
   \label{continuity_dim}
\end{equation}
\begin{equation}
   \frac{\partial {\bf u}}{\partial t} + \nabla \cdot ({\bf u} \otimes {\bf u}) 
   = \frac{1}{\rho} \left[ 
   - \nabla p + \mu \nabla^2 {\bf u} 
   + {\bf J} \times {\bf B} \right],
   \label{navier-stokes_dim}
\end{equation}
\begin{equation}
   \nabla \cdot {\bf B} = 0,
   \label{divergence_magnetic_dim}
\end{equation}
\begin{equation}
   \frac{\partial {\bf B}}{\partial t} 
   + \nabla \cdot \left( {\bf u} \otimes {\bf B}  - {\bf B} \otimes {\bf u} \right) 
   = \nabla \cdot \left( \nu_m \nabla {\bf B} \right),
   \label{magnetic_field_dim}
\end{equation}
where $t$ is the time, 
${\bf u}=(u, v, w)$ is the fluid velocity vector at the coordinates ${\bf x}=(x, y, z)$, 
$p$ is the pressure, 
and $\rho$ and $\mu$ are the density and viscosity coefficient of the fluid, respectively. 
${\bf J}=(J_x, J_y, J_z)$ is the current density, 
${\bf B}=(B_x, B_y, B_z)$ is the magnetic flux density, 
and $\nu_m$ is the magnetic diffusivity. 
The term ${\bf J} \times {\bf B}$ on the right side of the momentum conservation equation 
(\ref{navier-stokes_dim}) represents the Lorentz force.

The current density is given using Ampere's law as:
\begin{equation}
   {\bf J} = \nabla \times \left( \frac{1}{\mu_m} {\bf B} \right),
   \label{Ampere_dim}
\end{equation}
where $\mu_m$ is the magnetic permeability.

The above fundamental equations are nonlinear, 
and an analytic solution can be obtained only in a restricted case. 
We explain an analytic solution for a three-dimensional decaying vortex 
in the following.

\section{Three-dimensional Taylor decaying vortex}

\subsection{Analytic solution without applied magnetic field}

\citet{Antuono_2020} obtained a periodic three-dimensional analytic solution 
shown below using the method of \citet{Ethier&Steinman_1994}. 
Subscripts 1 and 2 represent two solutions.
\begin{eqnarray}
   u_{1,2} &=& \alpha U \left[ 
     \sin(k x + \theta_{1.2}) \cos(k y + \phi_{1.2}) \sin(k z + \psi_{1.2}) 
   \right. \nonumber \\
   && \quad \left. 
   - \cos(k z + \theta_{1.2}) \sin(k x + \phi_{1.2}) \sin(k y + \psi_{1.2}) 
   \right] e^{-3 \nu k^2 t}, \\
   v_{1,2} &=& \alpha U \left[ 
     \sin(k y + \theta_{1.2}) \cos(k z + \phi_{1.2}) \sin(k x + \psi_{1.2}) 
   \right. \nonumber \\
   && \quad \left. 
   - \cos(k x + \theta_{1.2}) \sin(k y + \phi_{1.2}) \sin(k z + \psi_{1.2}) 
   \right] e^{-3 \nu k^2 t}, \\
   w_{1,2} &=& \alpha U \left[ 
     \sin(k z + \theta_{1.2}) \cos(k x + \phi_{1.2}) \sin(k y + \psi_{1.2}) 
   \right. \nonumber \\
   && \quad \left. 
   - \cos(k y + \theta_{1.2}) \sin(k z + \phi_{1.2}) \sin(k x + \psi_{1.2}) 
   \right] e^{-3 \nu k^2 t}, \\
   p_{1,2} &=& - \rho \frac{|{\bf u}_{1,2}|^2}{2},
\end{eqnarray}
where $\alpha=4 \sqrt{2}/(3 \sqrt{3})$. 
$k$ is the wavenumber and $\lambda=2 \pi/k$ is the wavelength. 
The phase $\theta$, $\phi$, and $\psi$ are given as
\begin{equation}
   \theta_1 = \psi_1 - \frac{5 \pi}{6}, \quad 
   \phi_1 = \psi_1 - \frac{\pi}{6}, \quad 
   \psi_1 = \cos^{-1} \left( \frac{R}{\sqrt{1 + R^2}} \right),
\end{equation}
\begin{equation}
   \theta_2 = \phi_1, \quad 
   \phi_2 = \theta_1, \quad 
   \psi_2 = \psi_1,
\end{equation}
where $R$ is a parameter, 
and the value excluding the singularity value $R=\pm 1/\sqrt{3}$ is set. 
In this study, $R=0$ is set as in the existing study \citet{Antuono_2020}. 
The two solutions show similar distributions. 
The variables in the fundamental equations are non-dimensionalized 
using the reference values of length $\lambda$, velocity $U$, time $\lambda/U$, 
and pressure $\rho U^2$. 
The dimensionless exact solution is as follows:
\begin{eqnarray}
   u_{1,2} &=& \alpha \left[ 
     \sin(k x + \theta_{1.2}) \cos(k y + \phi_{1.2}) \sin(k z + \psi_{1.2}) 
   \right. \nonumber \\
   && \, \left. 
   - \cos(k z + \theta_{1.2}) \sin(k x + \phi_{1.2}) \sin(k y + \psi_{1.2}) 
   \right] e^{-3 \frac{k^2}{Re} t}, \\
   v_{1,2} &=& \alpha \left[ 
     \sin(k y + \theta_{1.2}) \cos(k z + \phi_{1.2}) \sin(k x + \psi_{1.2}) 
   \right. \nonumber \\
   && \, \left. 
   - \cos(k x + \theta_{1.2}) \sin(k y + \phi_{1.2}) \sin(k z + \psi_{1.2}) 
   \right] e^{-3 \frac{k^2}{Re} t}, \\
   w_{1,2} &=& \alpha \left[ 
     \sin(k z + \theta_{1.2}) \cos(k x + \phi_{1.2}) \sin(k y + \psi_{1.2}) 
   \right. \nonumber \\
   && \, \left. 
   - \cos(k y + \theta_{1.2}) \sin(k z + \phi_{1.2}) \sin(k x + \psi_{1.2}) 
   \right] e^{-3 \frac{k^2}{Re} t}, \\
   p_{1,2} &=& - \frac{{\bf u}_{1,2}|^2}{2}.
   \label{3D_Taylor_vortex}
\end{eqnarray}
where $k=2\pi$ is the non-dimensionalized wavenumber. 
The Reynolds number is defined as $Re=UL/\nu$.

This flow of the three-dimensional Taylor decaying vortex 
belongs to the Beltrami flow. 
Therefore, the velocity vector ${\bf u}$ and the vorticity vector ${\bf \omega}$ are parallel, 
and ${\bf u} \times {\bf \omega} = {\bf u} \times (\nabla \times {\bf u}) =0$. 
In addition, in the velocity and vorticity fields of this Beltrami flow, 
because the relationship of ${\bf \omega}_{1,2}=\pm \sqrt{3} k {\bf u}_{1,2}$ holds, 
the pressure is associated with the kinetic energy $\rho {\bf u} \cdot {\bf u}/2$ 
and enstrophy ${\bf \omega} \cdot {\bf \omega}/2$, respectively. 
Stagnation point is where the pressure is $p=0$. 
\citet{Antuono_2020} clarified that the high-pressure region 
including the stagnation point has a Y-shaped structure. 
The stagnation point occurs at the next position.
\begin{equation}
   (x_l, y_m, z_n) = - \frac{\psi}{k} (1, 1, 1) - \frac{\pi}{k} (l, m, n),
\end{equation}
where $l$, $m$, and $n$ are integers.

As in the two-dimensional case \citep{Taylor_1923}, 
the three-dimensional Taylor decaying vortex flow is also periodic, 
so the volume integral of the velocity is zero. 
Since the total amount of pressure corresponds to the volume integral 
of kinetic energy, it is a constant value of $-1/2$ for inviscid fluids. 
The volume-integrated kinetic energy is explained later.
\begin{equation}
   \int_V u dV = 0, \quad \int_V v dV = 0, \quad \int_V w dV = 0, \quad 
   \int_V p dV = - \frac{1}{2},
\end{equation}
where $V=1 \times 1 \times 1$.

The kinetic energy $K$ is given as
\begin{equation}
   K = \frac{1}{2} | {\bf u} |^2.
\end{equation}
The volume-integrated kinetic energy $\langle K \rangle$ 
and the average kinetic energy $K_{av}$ are given as
\begin{equation}
   \langle K \rangle = \int_V K dV = \frac{1}{2} e^{-\frac{6 k^2}{Re}t},
\end{equation}
\begin{equation}
   K_{av} = \frac{1}{V} \langle K \rangle 
   = \frac{1}{2} e^{-\frac{6 k^2}{Re}t},
\end{equation}
The time when the average kinetic energy $K_{av}$ is halved is $t=Re \ln(2)/(6k^2)$, 
and the time when $K_{av}$ decays to 10 \% of the initial kinetic energy 
is $t=Re \ln(10)/(6k^2)$.

\subsection{Analytic solution under applied magnetic field}

Assuming that the velocity vector and magnetic flux density vector represent 
the Beltrami flow, the exact solution of the velocity and pressure 
of the three-dimensional Taylor decaying vortex is expressed 
by the equation (\ref{3D_Taylor_vortex}) even under an applied magnetic field. 
Using the three-dimensional Taylor decaying vortex model 
proposed by \citet{Antuono_2020}, 
the magnetic flux density in an electromagnetic fluid 
under an applied magnetic field is given as
\begin{eqnarray}
   B_{x 1,2} &=& \alpha B \left[ 
     \sin(k x + \theta_{1.2}) \cos(k y + \phi_{1.2}) \sin(k z + \psi_{1.2}) 
   \right. \nonumber \\
   && \quad \left. 
   - \cos(k z + \theta_{1.2}) \sin(k x + \phi_{1.2}) \sin(k y + \psi_{1.2}) 
   \right] e^{-3 \nu_m k^2 t}, \\
   B_{y 1,2} &=& \alpha B \left[ 
     \sin(k y + \theta_{1.2}) \cos(k z + \phi_{1.2}) \sin(k x + \psi_{1.2}) 
   \right. \nonumber \\
   && \quad \left. 
   - \cos(k x + \theta_{1.2}) \sin(k y + \phi_{1.2}) \sin(k z + \psi_{1.2}) 
   \right] e^{-3 \nu_m k^2 t}, \\
   B_{z 1,2} &=& \alpha B \left[ 
     \sin(k z + \theta_{1.2}) \cos(k x + \phi_{1.2}) \sin(k y + \psi_{1.2}) 
   \right. \nonumber \\
   && \quad \left. 
   - \cos(k y + \theta_{1.2}) \sin(k z + \phi_{1.2}) \sin(k x + \psi_{1.2}) 
   \right] e^{-3 \nu_m k^2 t}.
\end{eqnarray}
The above equations satisfy the basic equations (\ref{divergence_magnetic_dim}) 
and (\ref{magnetic_field_dim}). 
As reference values dimensionless variables, 
the length is $\lambda$, velocity is $U$, time is $\lambda/U$, 
pressure is $\rho U^2$, and magnetic density is $B$. 
The exact solution is non-dimensionalized as follows:
\begin{eqnarray}
   B_{x 1,2} &=& \alpha \left[ 
     \sin(k x + \theta_{1.2}) \cos(k y + \phi_{1.2}) \sin(k z + \psi_{1.2}) 
   \right. \nonumber \\
   && \, \left. 
   - \cos(k z + \theta_{1.2}) \sin(k x + \phi_{1.2}) \sin(k y + \psi_{1.2}) 
   \right] e^{-3 \frac{k^2}{Re_m} t}, \\
   B_{y 1,2} &=& \alpha \left[ 
     \sin(k y + \theta_{1.2}) \cos(k z + \phi_{1.2}) \sin(k x + \psi_{1.2}) 
   \right. \nonumber \\
   && \, \left. 
   - \cos(k x + \theta_{1.2}) \sin(k y + \phi_{1.2}) \sin(k z + \psi_{1.2}) 
   \right] e^{-3 \frac{k^2}{Re_m} t}, \\
   B_{z 1,2} &=& \alpha \left[ 
     \sin(k z + \theta_{1.2}) \cos(k x + \phi_{1.2}) \sin(k y + \psi_{1.2}) 
   \right. \nonumber \\
   && \, \left. 
   - \cos(k y + \theta_{1.2}) \sin(k z + \phi_{1.2}) \sin(k x + \psi_{1.2}) 
   \right] e^{-3 \frac{k^2}{Re_m} t},
   \label{3D_Taylor_vortex_mag}
\end{eqnarray}
where $k=2\pi$ is the non-dimensionalized wavenumber. 
The magnetic Reynolds number is defined as $Re_m=UL/\nu_m$.

Since this three-dimensional magnetic flux density flow is the same 
as the Beltrami flow, 
the magnetic flux density vector ${\bf B}$ and the current density vector ${\bf J}$ are parallel, 
and ${\bf B} \times {\bf J} = {\bf B} \times (\nabla \times {\bf B}) = 0$. 
Therefore, the Lorentz force does not act, and work does not occur. 
Assuming that the kinematic viscosity and magnetic diffusivity are zero, 
the kinetic and magnetic energies are conserved in a periodic flow, 
and the total energy is also conserved. 
Therefore, it is useful to use this decaying vortex model as a benchmark test problem 
for verifying the validity of the energy conservation characteristics 
in a calculation method.

Since the flow of the three-dimensional Taylor decaying vortex is periodic, 
the volume integral of the magnetic flux density is zero like the velocity.
\begin{equation}
   \int_V B_x dV = 0, \quad \int_V B_y dV = 0, \quad \int_V B_z dV = 0,
\end{equation}
where $V=1 \times 1 \times 1$.

The magnetic energy $M$ is given as
\begin{equation}
   M = \frac{1}{2 Al^2} | {\bf B} |^2.
\end{equation}
This magnetic energy $M$ corresponds to dimensionless magnetic pressure. 
The volume-integrated magnetic energy $\langle M \rangle$ 
and the average magnetic energy $M_{av}$ are obtained as follows:
\begin{equation}
   \langle M \rangle = \int_V M dV = \frac{1}{2 Al^2} e^{-\frac{6 k^2}{Re_m}t},
\end{equation}
\begin{equation}
   M_{av} = \frac{1}{V} \langle M \rangle 
   = \frac{1}{2 Al^2} e^{-\frac{6 k^2}{Re_m}t},
\end{equation}
The total energy $E_t$ is $E_t=K + M$. 
The volume-integrated total energy $\langle E_t \rangle$ 
and the average total energy $E_{t av}$ are obtained as follows:
\begin{equation}
   \langle E_t \rangle 
   = \frac{1}{2} \left( e^{-\frac{6 k^2}{Re}t} 
       + \frac{1}{Al^2} e^{-\frac{6 k^2}{Re_m}t} \right),
\end{equation}
\begin{equation}
   E_{t av} = K_{av} + M_{av} 
   = \frac{1}{2} \left( e^{-\frac{6 k^2}{Re}t} 
       + \frac{1}{Al^2} e^{-\frac{6 k^2}{Re_m}t} \right).
\end{equation}
Cross helicity $H_c$ is given as
\begin{equation}
   H_c = \frac{1}{Al} {\bf u} \cdot {\bf B}.
\end{equation}
The volume-integrated cross helicity $\langle H_c \rangle$ 
and the average cross helicity $H_{c av}$ are obtained as follows:
\begin{equation}
   \langle H_c \rangle = \int_V H_c dV 
   = \frac{1}{Al} e^{-\frac{3 k^2}{Re}t} e^{-\frac{3 k^2 \pi^2}{Re_m}t},
\end{equation}
\begin{equation}
   H_{c av} = \frac{1}{V} \langle H_c \rangle 
   = \frac{1}{Al} e^{-\frac{3 k^2}{Re}t} e^{-\frac{3 k^2 \pi^2}{Re_m}t}.
\end{equation}

\section{Numerical method and calculation conditions}

\subsection{Numerical method}

For stable calculations, transport quantities such as kinetic energy 
must be conserved discretely in periodic flows of inviscid fluids. 
In addition, the analytical characteristics of the governing equation 
must hold discretely, 
and the compatibility between the conservative and non-conservative forms 
of convection term must be satisfied discretely. 
Similar to the previous research \citep{Ham_et_al_2002, Morinishi_2009}, 
we use the implicit midpoint law to discretize the time derivatives 
in the governing equations and perform time marching. 
The second-order central difference scheme is used 
for the discretization of the space derivatives. 
The simplified marker and cell method \citep{Amsden&Harlow_1970} is applied 
to solve the discretized equations. 
In this study, the Lorentz force is discretized using compact interpolation 
\citep{Yanaoka_2022}, 
which can maintain the compatibility between conservative and non-conservative forms 
of the Lorentz force. 
We describe the Lorentz force discretization method 
in Appendix \ref{subsection_Lorentz}.

\subsection{Calculation conditions for three-dimensional Taylor decaying vortex}

The calculation area is a cube with one side $\lambda$. 
A uniform grid of $41^3$ is used for the calculation. 
For error evaluation and grid dependency verification, 
we also use the grids with dimensions of $11^3$, $21^3$, and $81^3$. 
Exact solutions ${\bf u}_1$ and $p_1$ are given as initial conditions, 
and periodic boundary conditions are set as boundary conditions.

First, we verify the validity and stability of the present calculation method. 
To confirm whether non-physical kinetic energy is generated, 
we analyse the inviscid flow at $Re=\infty$. 
The calculation is performed up to the time $t/(L/U)=10$ under the condition 
that the Courant number is CFL=0.5.

Regarding the calculation conditions of the decaying vortex, 
the Reynolds numbers are $Re=10$, 50, $10^2$, and $10^3$. 
We calculate until the average kinetic energy is half 
and compare the calculation result with the exact solution. 
For each Reynolds number, the time step is given 
so that the Courant number CFL is constant regardless of the grid. 
The Courant number for Reynolds numbers other than $Re=10$ is CFL=0.25. 
At $Re=10$, the time until the average kinetic energy is half is short, 
so the Courant number was set to CFL=0.05.

Similar to the earlier study \citep{Antuono_2020}, 
this study performs a long-time calculation 
to confirm whether the transition to turbulent flow is observed. 
As in the previous study, the Reynolds number is $Re=10^3$. 
In this calculation, we also use a grid finer than the number of grid points 
used in the existing research. 
The Courant number is set to CFL=0.1 and 0.5 
to verify the influence of time increments when the number of grid points is 41. 
In other grids, the Courant number is CFL=0.5. 
In addition, we investigate the effect of initial velocity disturbance 
on the flow transition. 
Similar to the previous study\citep{Morinishi_2009}, 
we use the vector potential with uniform random numbers to generate 
an initial disturbance that satisfies the continuity equation. 
The disturbance levels are values corresponding to 1
of the initial average kinetic energy. 
As an initial condition, we use the value obtained by adding this initial disturbance 
to the exact solution of velocity.

\subsection{Calculation conditions for three-dimensional Taylor decaying vortex 
under applied magnetic field}

Next, we consider the analysis of a three-dimensional Taylor decaying vortex 
under an applied magnetic field. 
The grids and boundary conditions used are similar to the analytical model 
described in the previous subsection. 
The exact solutions ${\bf u}_1$, $p_1$, and ${\bf B}_1$ are given as initial conditions.

Regarding the calculation conditions, 
the Reynolds numbers are set to $Re=10^2$ and $10^4$, 
referring to the existing research on two-dimensional Taylor decaying vortex 
\citep{Liu&Wang_2001, Yanaoka_2022}. 
The magnetic Reynolds numbers are $Re_m=1$ and 50, 
and the Alfv\'{e}n number is $Al=1$. 
We calculate under the condition that the Courant number CFL=0.25.

\begin{figure}[!t]
\begin{minipage}{0.48\linewidth}
\begin{center}
\includegraphics[trim=0mm 0mm 0mm 0mm, clip, width=70mm]{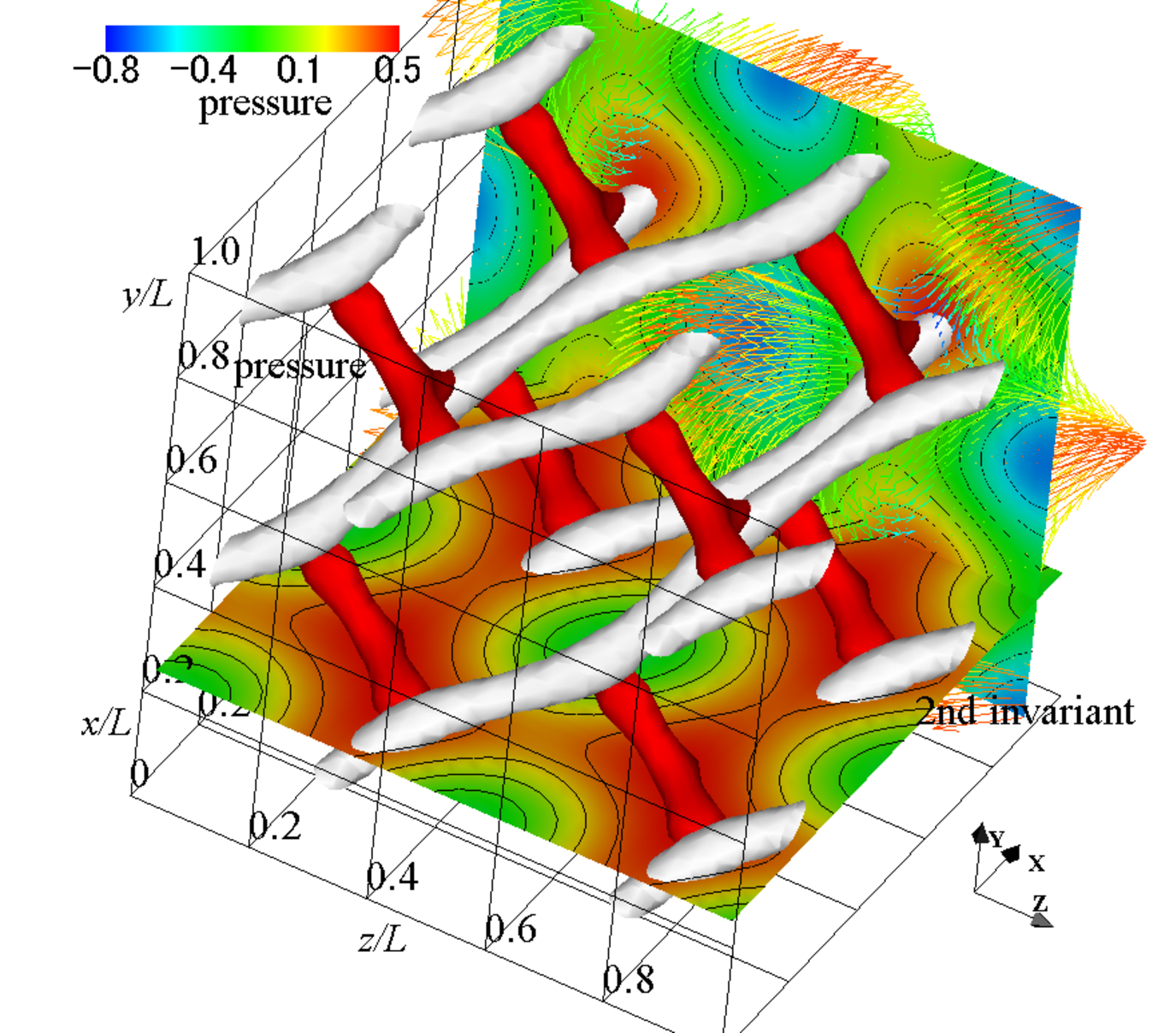} \\
(a) velocity vectors, pressure contour, isosurface of pressure, 
and isosurface of 2nd invariant of velocity gradient tensor
\end{center}
\end{minipage}
\hspace{0.02\linewidth}
\begin{minipage}{0.48\linewidth}
\begin{center}
\includegraphics[trim=0mm 0mm 0mm 0mm, clip, width=70mm]{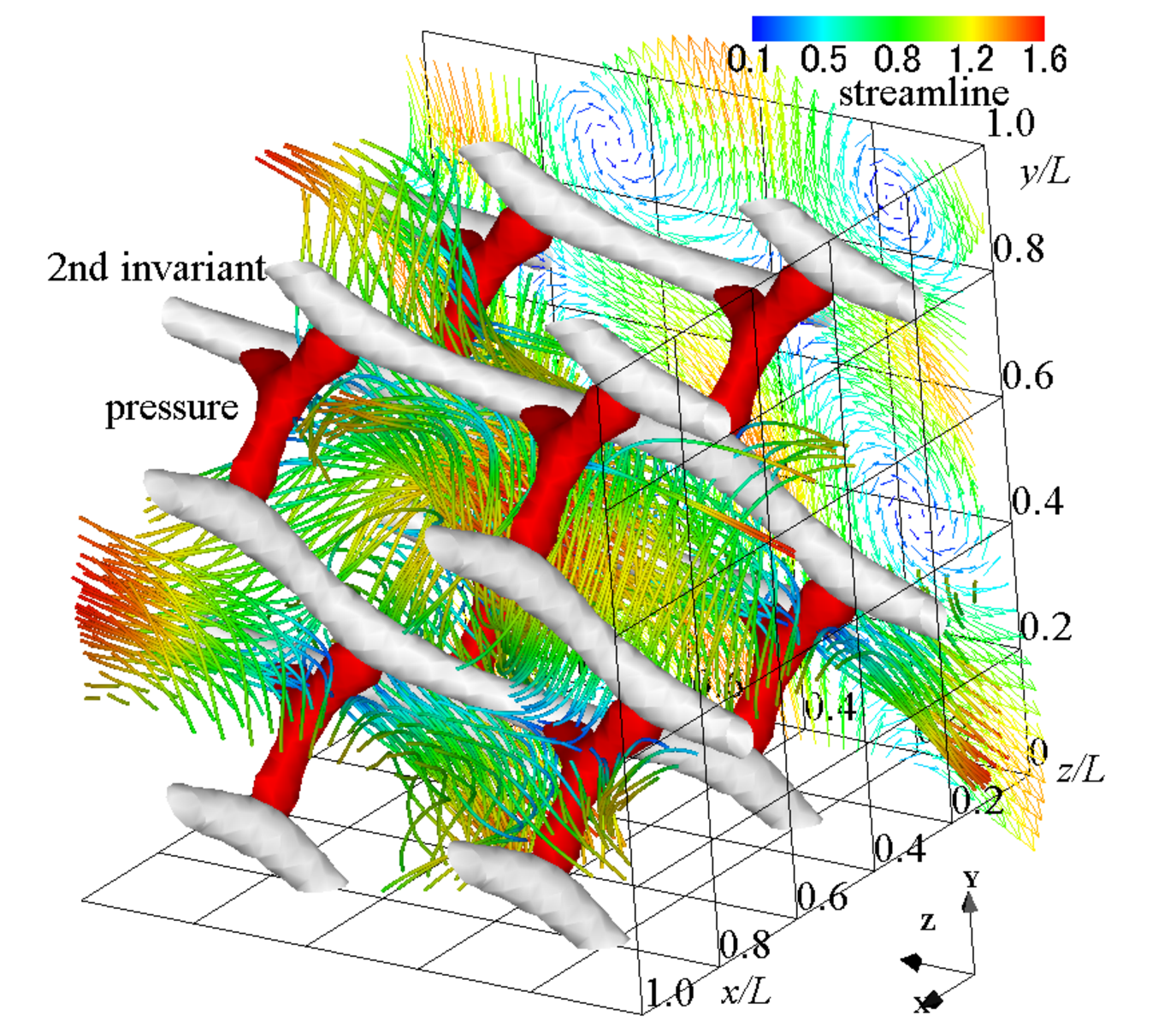} \\
(b) velocity vectors, streamlines, isosurface of pressure, 
and isosurface of 2nd invariant of velocity gradient tensor \\
\vspace*{1.5\baselineskip}
\end{center}
\end{minipage}
\vspace*{0.5\baselineskip}
\begin{minipage}{0.48\linewidth}
\begin{center}
\includegraphics[trim=0mm 0mm 0mm 0mm, clip, width=70mm]{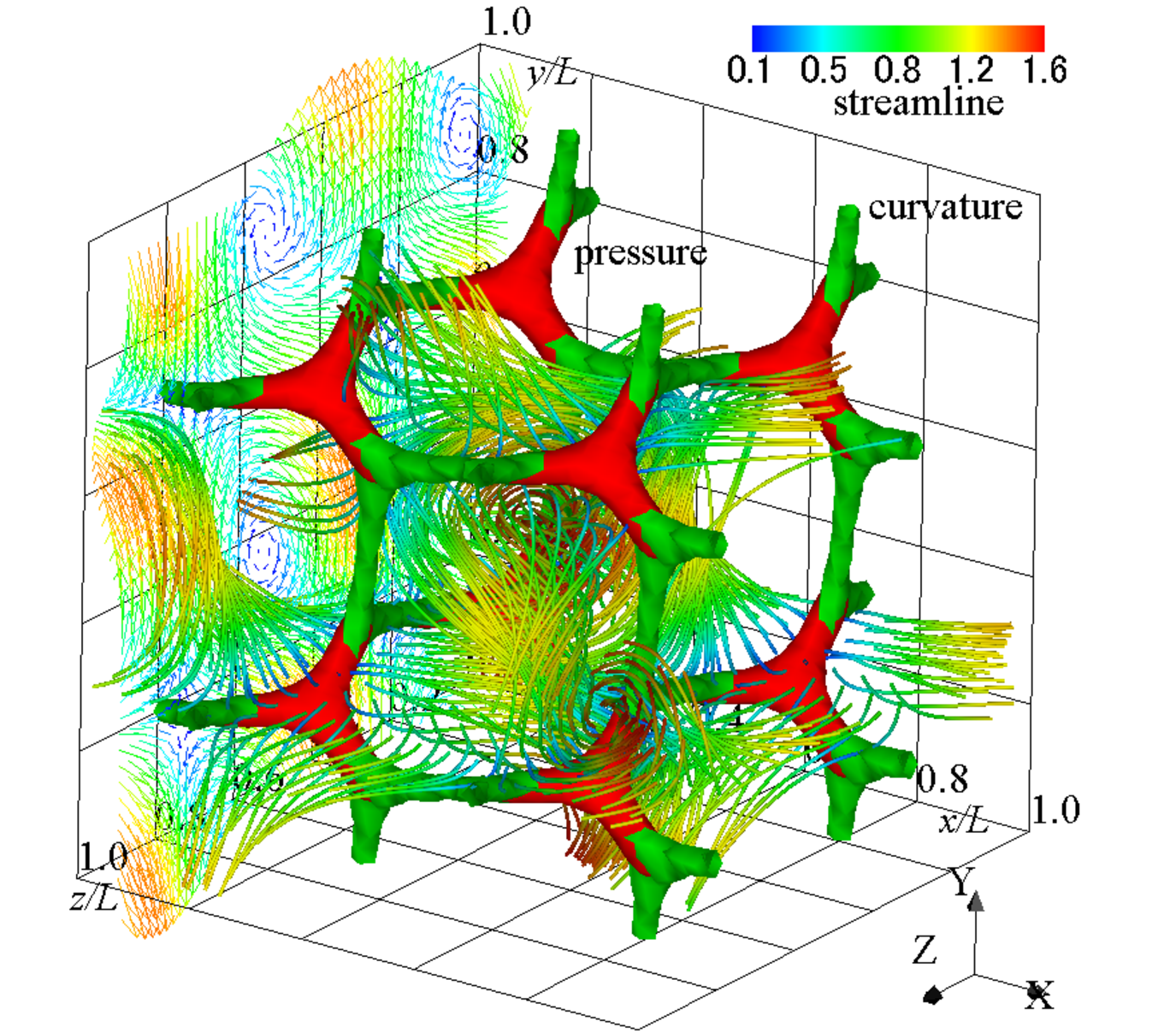} \\
(c) velocity vectors, streamlines, isosurface of pressure, 
and isosurface of the curvature of equipressure surface
\end{center}
\end{minipage}
\begin{minipage}{0.48\linewidth}
\begin{center}
\includegraphics[trim=0mm 0mm 0mm 0mm, clip, width=70mm]{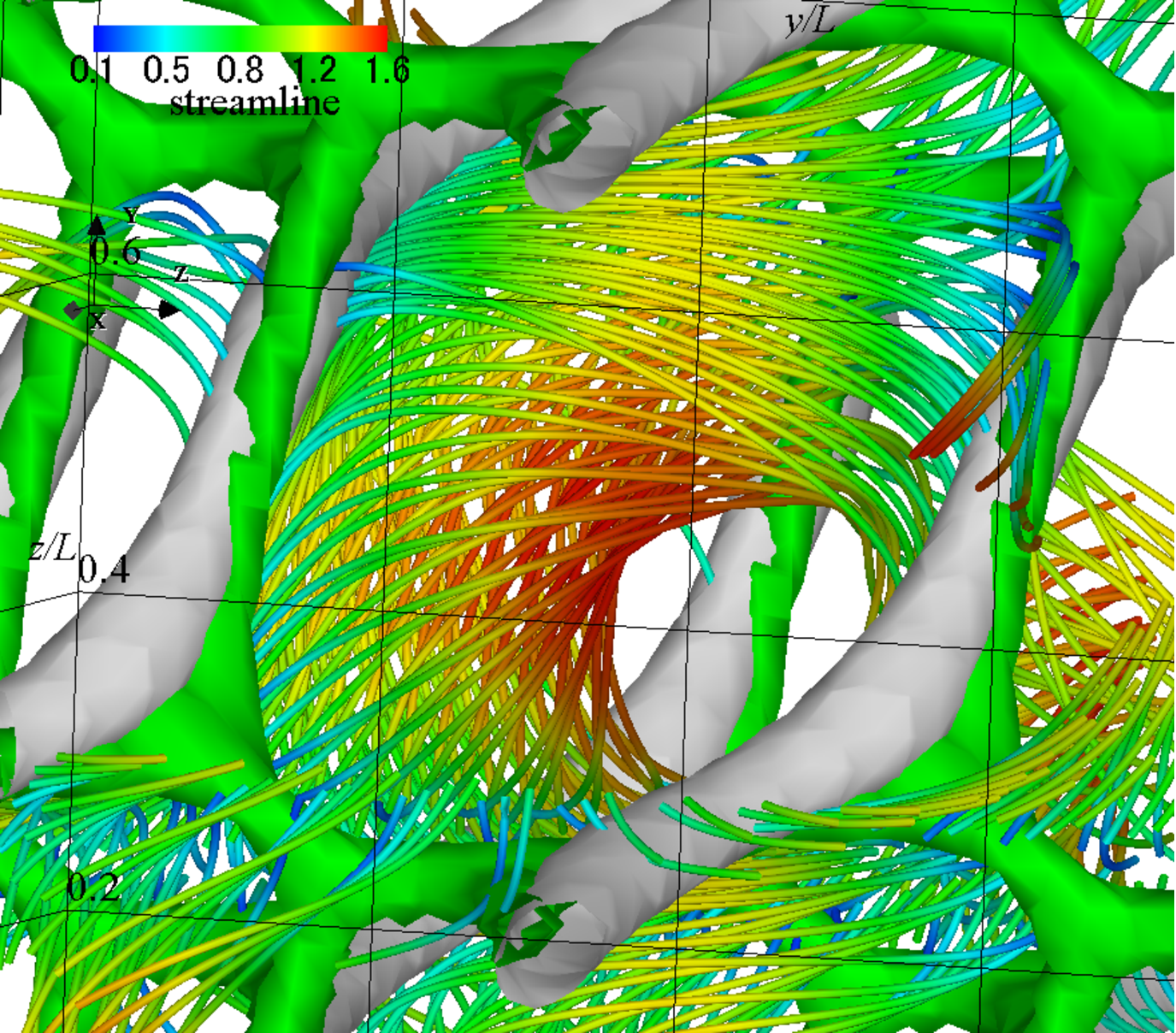} \\
(d) streamlines, isosurface of 2nd invariant of velocity gradient tensor, 
and isosurface of the curvature of equipressure surface
\end{center}
\end{minipage}
%
\caption{Velocity vectors, streamlines, pressure contour, 
and various isosurfaces for analytic solution: 
The red, silver and green isosurfaces show the pressure, 
2nd invariant of the velocity gradient tensor, 
and curvature of equipressure surface, respectively.}
\label{analytical_solution}
\end{figure}

\subsection{Extraction of pressure structure}

We explain the method of extracting low- or high-pressure regions. 
If pressure distribution is concentric around a vortex tube, 
we can identify the vortex tube by displaying the isosurface of the pressure. 
However, when the vortex tube and shear layer coexist, 
the pressure changes due to the two structures, 
so we cannot extract only the vortex tube. 
Because the radius of a thin vortex tube is small, 
we can identify the thin vortex tube by visualising a vortex tube 
with a large curvature. 
Therefore, by calculating the curvature of an equipressure surface 
and displaying the isosurface with a large curvature, 
we can identify a vortex tube with a small radius of curvature. 
Now we consider the case where pressure is high at the centre of a concentric circle 
and low at the periphery. 
The curvature of the pressure isosurface can be defined as follows:
\begin{equation}
   \kappa_p = - \nabla \cdot \hat{{\bf n}}, \quad 
   \hat{{\bf n}} = \frac{{\bf n}}{|{\bf n}|} = \frac{\nabla p}{|\nabla p|},
\end{equation}
where $\hat{{\bf n}}$ is a unit normal vector on the isosurface. 
Therefore, we can visualise a high-pressure region 
by displaying the high value of $\kappa_p>0$. 
Conversely, we can extract a low-pressure region 
by showing the low value of $\kappa_p<0$. 
$1/|\kappa_p|$ is the radius of curvature of the vortex tube.

\subsection{Characteristics of analytic solutions}

Figure \ref{analytical_solution} shows the flow field of the analytic solution 
at $t=0$. 
The velocity vectors, streamlines, and the contour 
and isosurface of pressure are shown. 
The second invariant of velocity gradient tensor and the curvature 
of an equipressure surface are also displayed in an isosurface form. 
The red isosurface of pressure shows the distribution of dimensionless pressure $p=0.48$, 
and we can see the pressure field near a stagnation point. 
The second invariant $Q$ of the velocity gradient tensor is a quantity 
that expresses the magnitude relationship between the strain rate tensor 
and the vorticity tensor. 
Structures with $Q<0$ represent regions of high shear rates, 
where the high viscous dissipation rate of kinetic energy occurs. 
The second invariant of the velocity gradient tensor is a negative value of $Q=-52$ 
and represents a tubular high-shear region where the strain rate tensor increases. 
Along the direction of the vector $(1, 1, 1)$, the tubular structure exists 
a little away from the stagnation point. 
The high-pressure region with the stagnation point is a Y-shaped structure \citep{Antuono_2020}. 
To extract the structure of a high-pressure area near a stagnation point, 
we calculate the curvature of the isosurface of the pressure. 
The green isosurface represents its curvature, 
and the magnitude of the curvature is $\kappa_p=48$. 
The high-pressure regions, which indicate the low-velocity areas 
with stagnation points, are connected in a mesh pattern, 
and a distorted cube structure appears. 
We can see that the isosurface of curvature buries the isosurface of pressure, 
and the cube structure represents the structure of the pressure field. 
The pressure is low at the centre of this cube structure, 
and when looking at the streamline, a swirling flow occurs 
so as to go around the low-pressure region. 
The tubular high-shear structure passes through the high-pressure region, 
and the pressure at the centre of the tubular high-shear structure is higher than 
the central pressure of the decaying vortex. 
No clear rotational flow occurs around the tubular high-shear structure.

\section{Results and discussion}

\subsection{Analysis of three-dimensional Taylor decaying vortex}

\begin{figure}[!t]
\begin{minipage}{0.48\linewidth}
\begin{center}
\includegraphics[trim=0mm 0mm 0mm 0mm, clip, width=70mm]{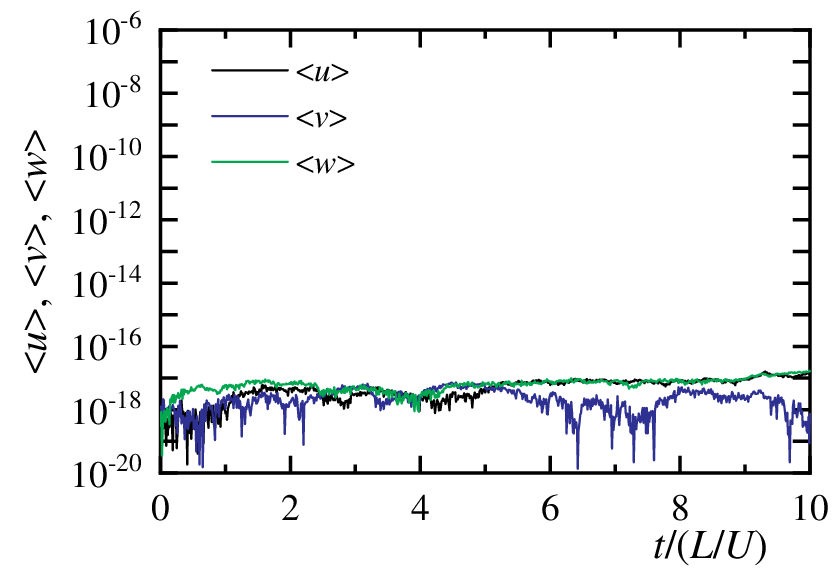} \\
(a) velocity
\end{center}
\end{minipage}
\hspace{0.02\linewidth}
\begin{minipage}{0.48\linewidth}
\begin{center}
\includegraphics[trim=0mm 0mm 0mm 0mm, clip, width=70mm]{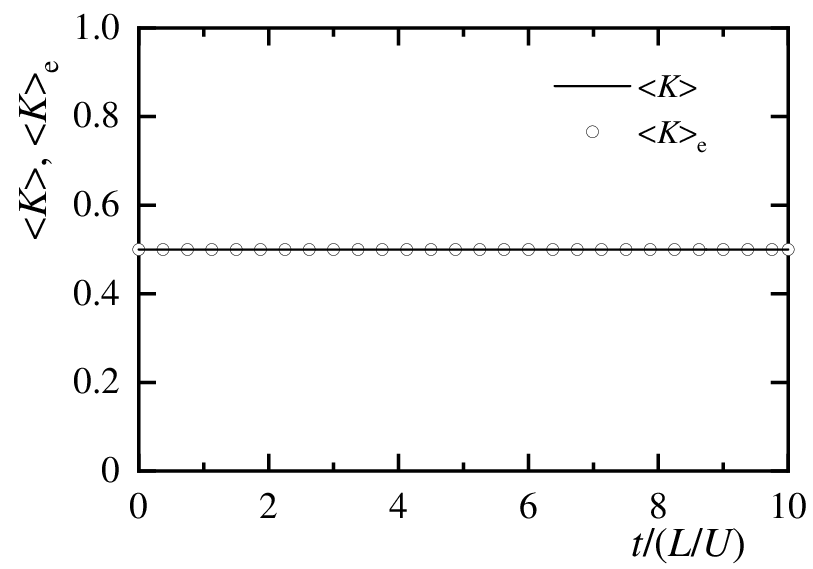} \\
(b) kinetic energy
\end{center}
\end{minipage}
%
\caption{Total amounts of velocities and kinetic energy 
for inviscid analysis.}
\label{sum_quantity_inviscid}
\end{figure}

Figure \ref{sum_quantity_inviscid} shows the time variation of the total amount 
of transport quantity in an inviscid analysis. 
$\langle u \rangle$, $\langle v \rangle$, and $\langle w \rangle$ 
are the volume integrals of velocities of $x$-, $y$-, and $z$-directions, respectively, 
and $\langle K \rangle$ are the volume integral of kinetic energy. 
$\langle K \rangle_e$ is the exact solution. 
Since the flow is periodic, each total amount of the transport quantity is preserved. 
The total amounts of the velocities change at the level of rounding error, 
indicating that the conservation characteristics of the present calculation method 
are excellent. 
In addition, the kinetic energy is kept constant and consistent with the exact solution, 
and no non-physical energy is generated. 
Therefore, since this calculation method has excellent stability 
and does not generate non-physical kinetic energy, 
it is considered that this method can simulate a decaying vortex in a viscous fluid 
without destroying its characteristics.

\begin{figure}[!t]
\begin{minipage}{0.48\linewidth}
\begin{center}
\includegraphics[trim=0mm 0mm 0mm 0mm, clip, width=70mm]{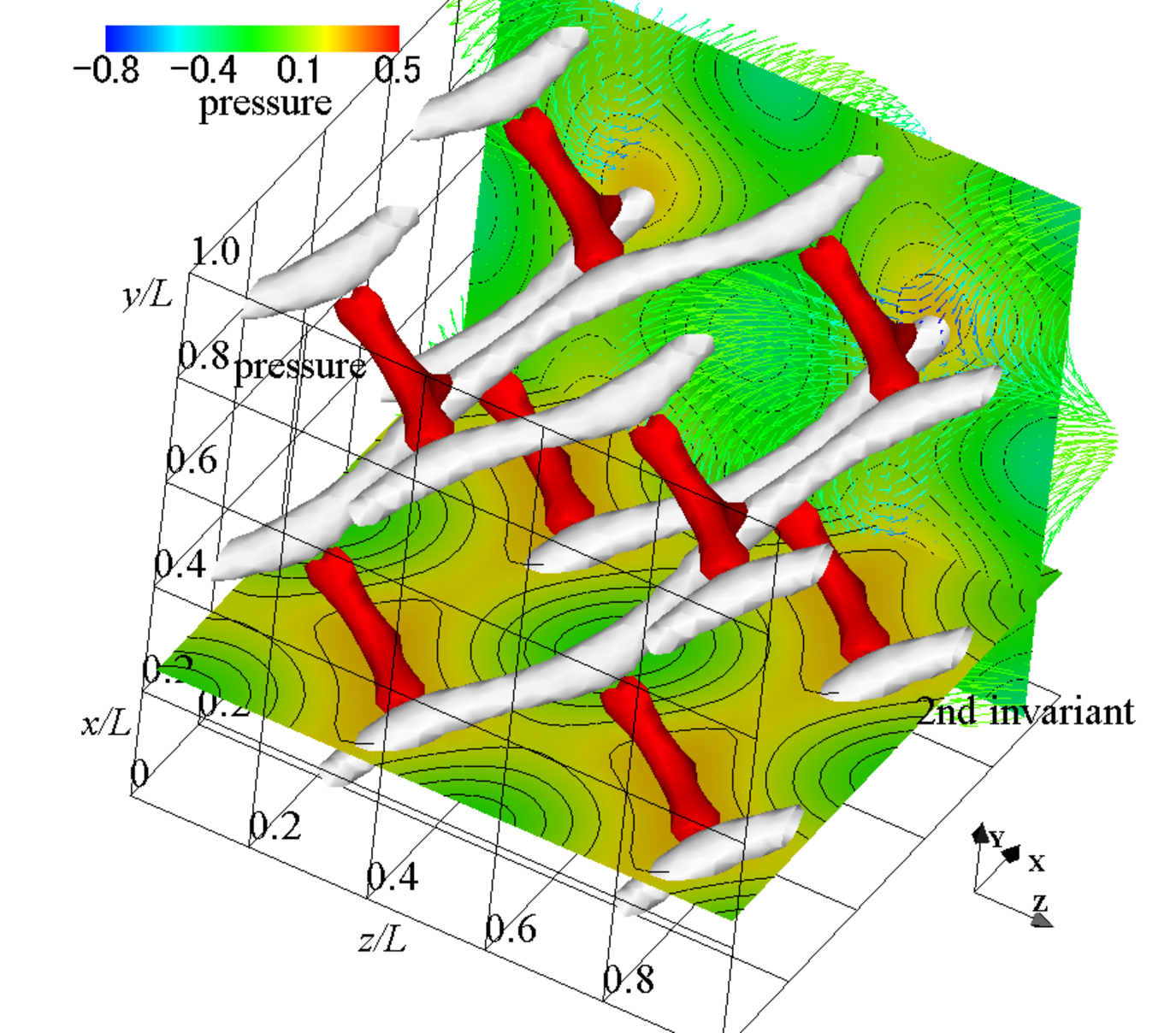} \\
(a) velocity vectors, pressure contour, isosurface of pressure, 
and isosurface of 2nd invariant of velocity gradient tensor
\end{center}
\end{minipage}
\hspace{0.02\linewidth}
\begin{minipage}{0.48\linewidth}
\begin{center}
\includegraphics[trim=0mm 0mm 0mm 0mm, clip, width=70mm]{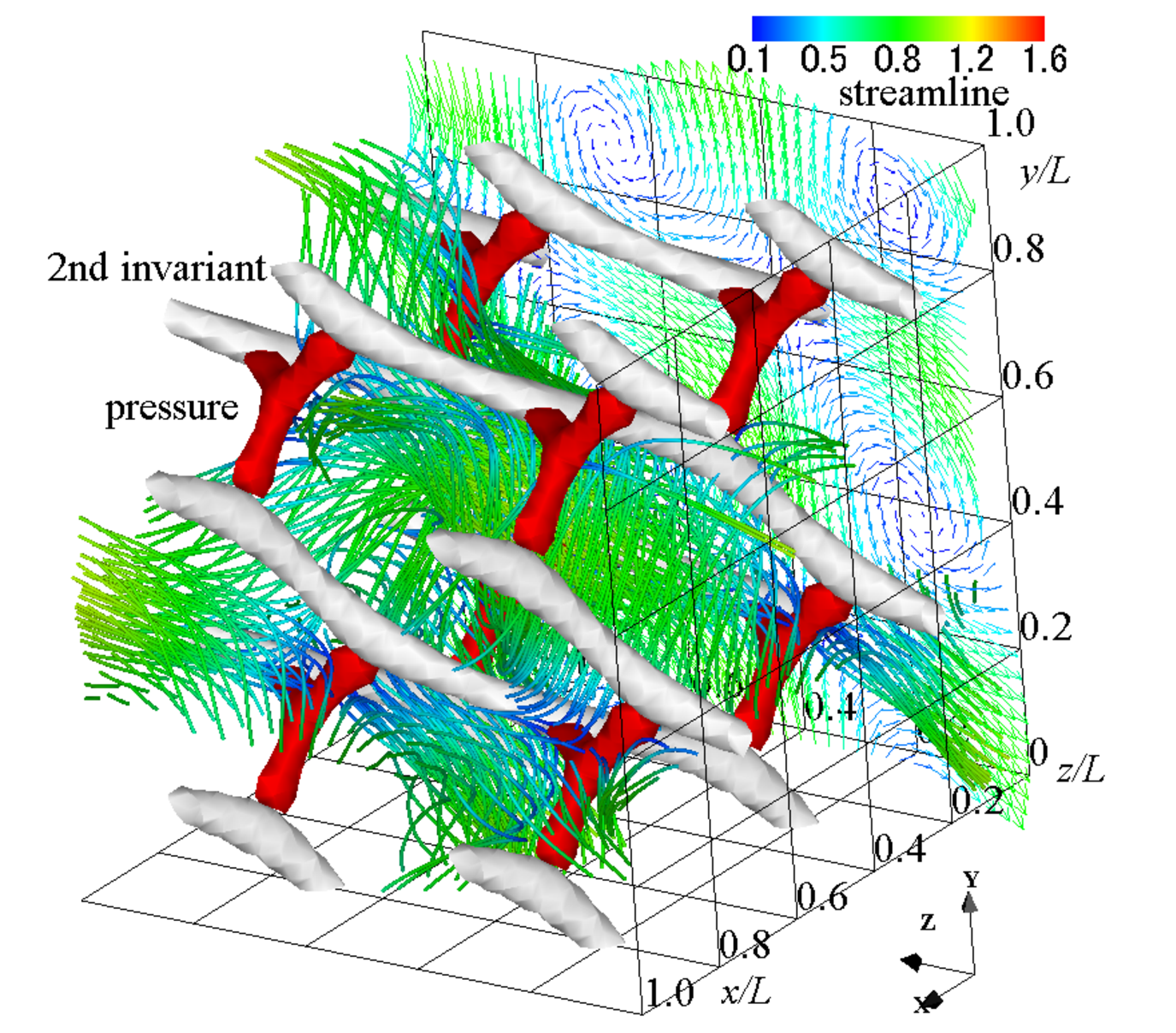} \\
(b) velocity vectors, streamlines, isosurface of pressure, 
and isosurface of 2nd invariant of velocity gradient tensor \\
\vspace*{1.5\baselineskip}
\end{center}
\end{minipage}
\vspace*{0.5\baselineskip}
\begin{minipage}{0.48\linewidth}
\begin{center}
\includegraphics[trim=0mm 0mm 0mm 0mm, clip, width=70mm]{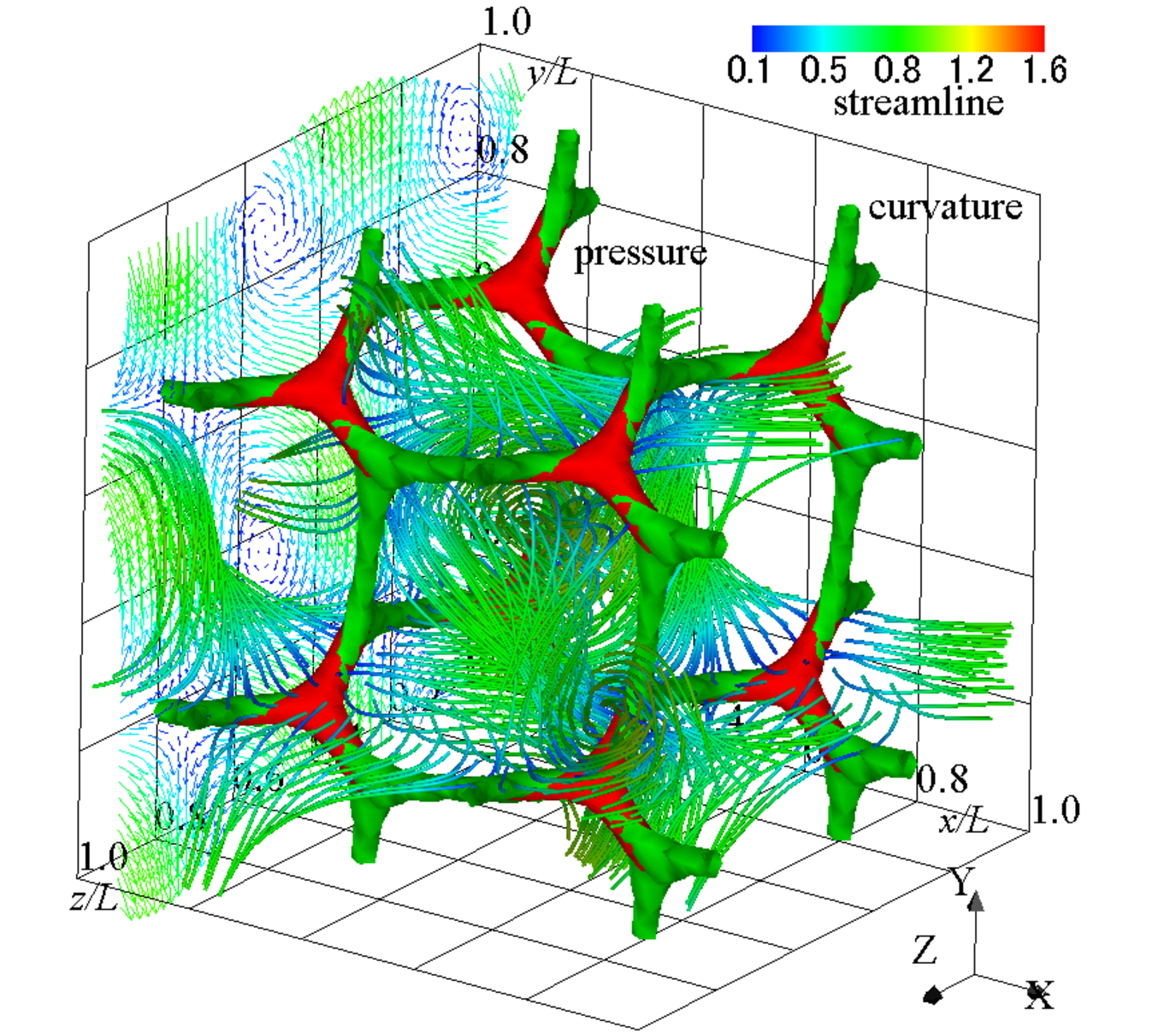} \\
(c) velocity vectors, streamlines, isosurface of pressure, 
and isosurface of the curvature of equipressure surface
\end{center}
\end{minipage}
\begin{minipage}{0.48\linewidth}
\begin{center}
\includegraphics[trim=0mm 0mm 0mm 0mm, clip, width=70mm]{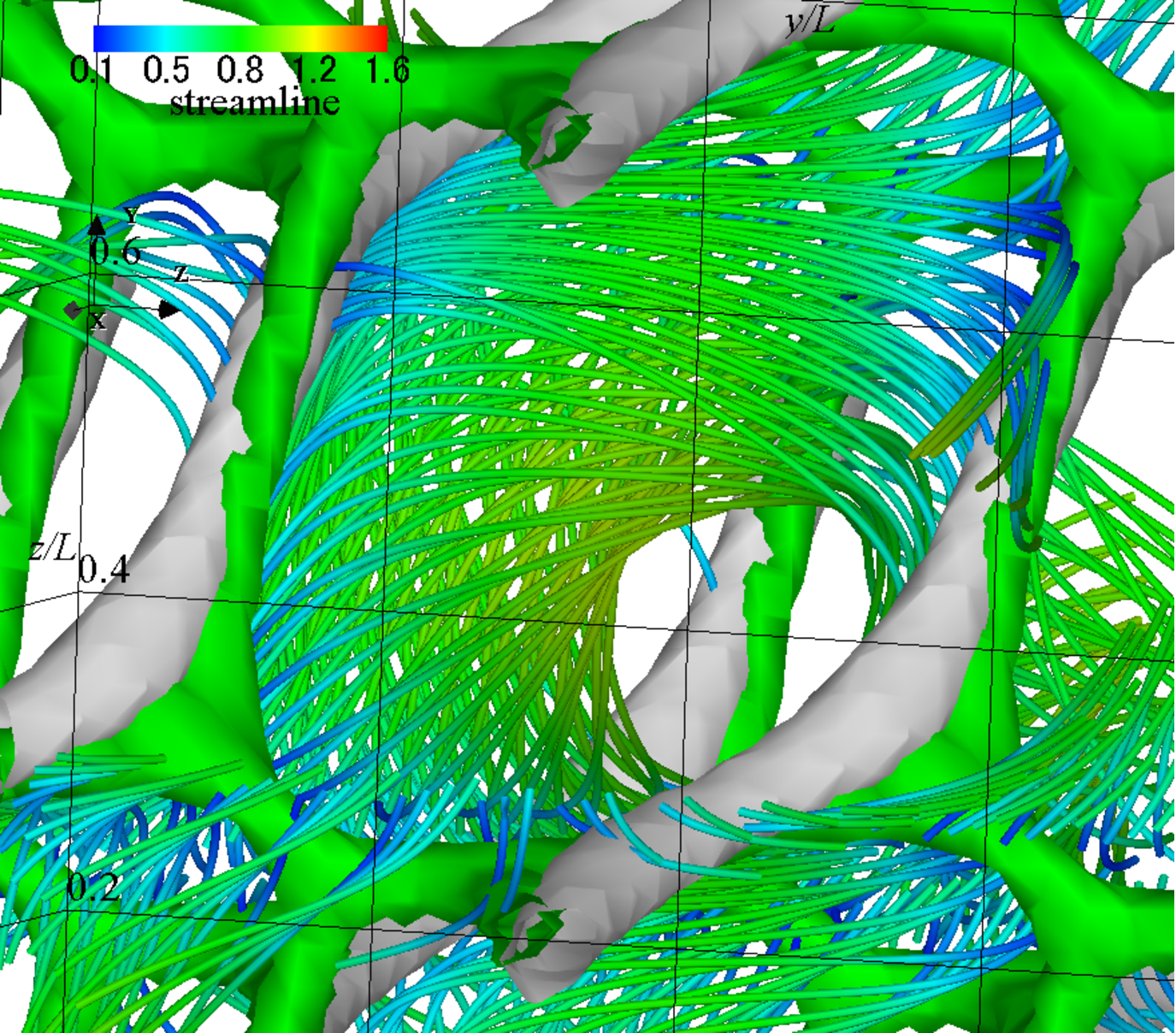} \\
(d) streamlines, isosurface of 2nd invariant of velocity gradient tensor, 
and isosurface of the curvature of equipressure surface
\end{center}
\end{minipage}
%
\caption{Velocity vectors, streamlines, pressure contour, 
and various isosurfaces: $Re=10^2$, $t/(L/U)=0.3$; 
The red, silver and green isosurfaces show the pressure, 
2nd invariant of the velocity gradient tensor, 
and curvature of equipressure surface, respectively.}
\label{flow_Re100}
\end{figure}

Figure \ref{flow_Re100} shows the flow field of $Re=10^2$ at time $t/(L/U)=0.3$. 
The isosurface values were set 
so that the size of the structure represented by the isosurface was the same 
as the structure size shown in figure \ref{analytical_solution}. 
The red isosurface expresses the distribution of the dimensionless pressure $p=0.24$. 
The pressure near the stagnation point becomes lower than the initial value 
shown in figure \ref{analytical_solution}. 
The second invariant of the velocity gradient tensor is $Q=-26$, 
and we find that the tubular high-shear structure with the initial strength of $Q=-52$ 
diffuses and decays. 
Since the vorticity of the Taylor decaying vortex decreases with time, 
the magnitude of the strain rate tensor inside a high-shear structure 
generated by the interaction between vortices also decreases. 
The curvature of the pressure isosurface is $\kappa_p=48$, 
which is the same as the initial value. 
Over time, the cube structure in which the high-pressure regions are connected 
in a mesh pattern is maintained. 
The velocity of the swirling flow inside this cubic structure becomes slow.

\begin{figure}[!t]
\begin{minipage}{0.48\linewidth}
\begin{center}
\includegraphics[trim=0mm 0mm 0mm 0mm, clip, width=70mm]{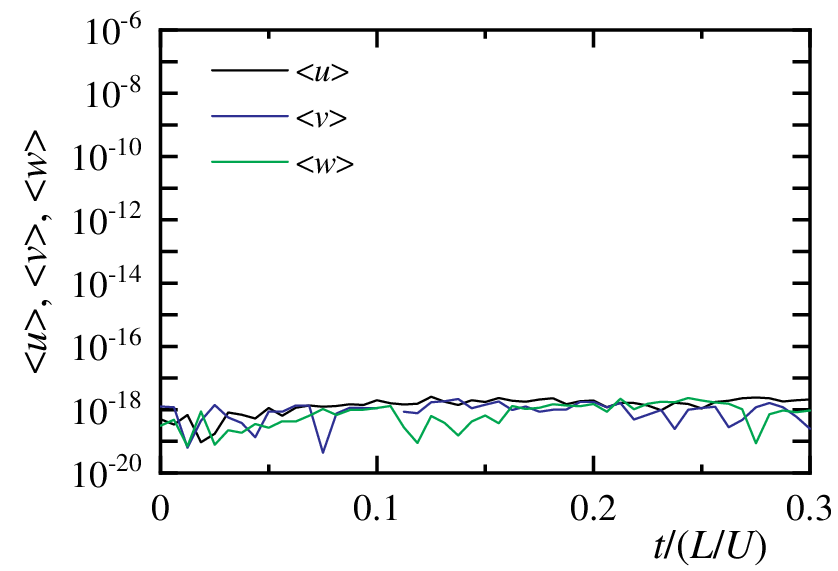} \\
(a) velocity
\end{center}
\end{minipage}
\hspace{0.02\linewidth}
\begin{minipage}{0.48\linewidth}
\begin{center}
\includegraphics[trim=0mm 0mm 0mm 0mm, clip, width=70mm]{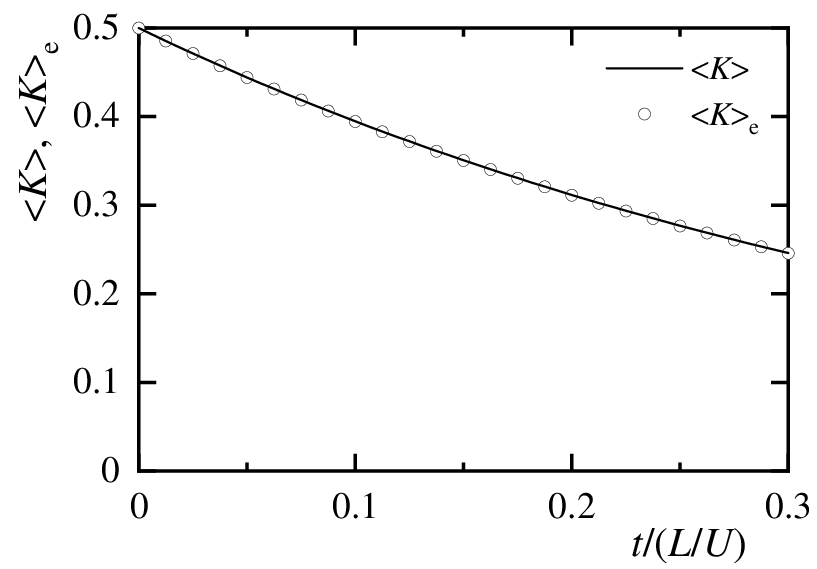} \\
(b) kinetic energy
\end{center}
\end{minipage}
%
\caption{Total amounts of velocities and kinetic energy at $Re=10^2$.}
\label{sum_quantity_Re100}
\end{figure}

\begin{figure}[!t]
\begin{center}
\includegraphics[trim=0mm 0mm 0mm 0mm, clip, width=70mm]{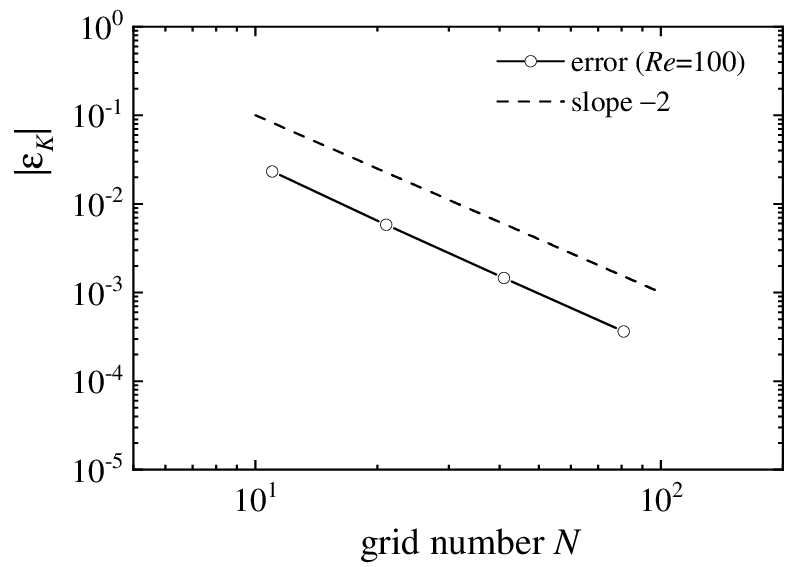} \\
\end{center}
\vspace*{-1.0\baselineskip}
\caption{Error of kinetic energy: $Re=10^2$, $t/(L/U)=0.3$.}
\label{error_K_Re100}
\end{figure}

Figure \ref{sum_quantity_Re100} shows the time variations of the total amounts 
of velocities and kinetic energy at $Re=10^2$. 
The total amount of each velocity is the level of rounding error. 
The kinetic energy decays over time. 
This calculation results well agree with the exact solution, 
indicating that this calculation method accurately captures the characteristics 
of the decaying vortex. 
The three-dimensional periodic decaying vortex proposed by \citet{Antuono_2020} 
is an effective model for investigating the conservation characteristics 
of calculation methods and changes in kinetic energy.

\begin{figure}[!t]
\begin{minipage}{0.48\linewidth}
\begin{center}
\includegraphics[trim=0mm 0mm 0mm 0mm, clip, width=70mm]{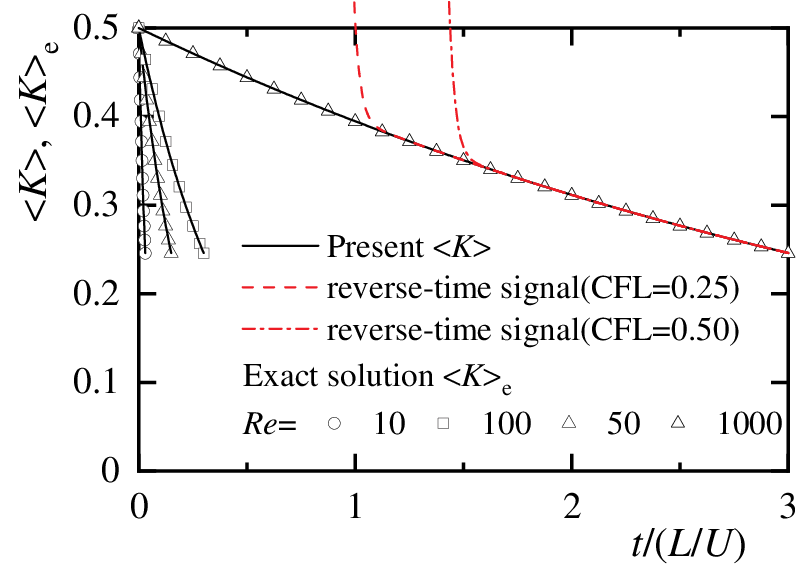} \\
(a) kinetic energy
\end{center}
\end{minipage}
\hspace{0.02\linewidth}
\begin{minipage}{0.48\linewidth}
\begin{center}
\includegraphics[trim=0mm 0mm 0mm 0mm, clip, width=70mm]{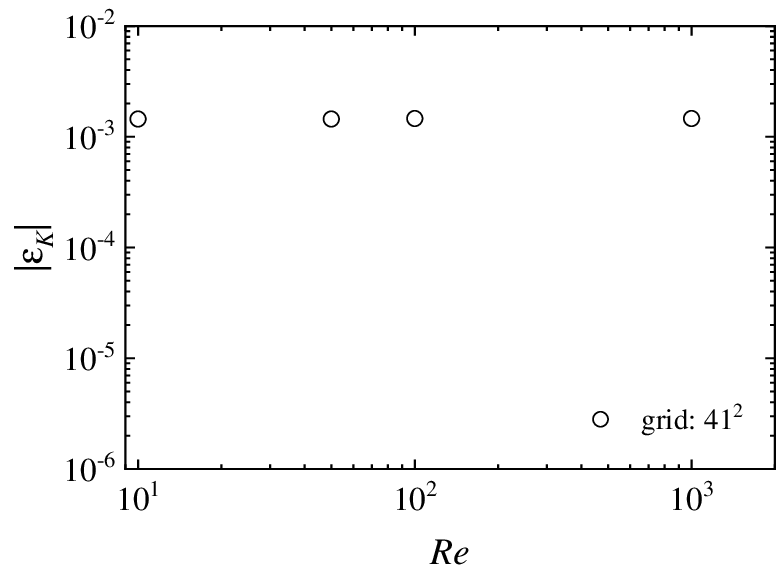} \\
(b) error
\end{center}
\end{minipage}
%
\caption{Total amount of kinetic energy, 
and error of kinetic energy: $Re=10$, 50, $10^2$, and $10^3$.}
\label{error_K_Re}
\end{figure}

\begin{figure}[!t]
\begin{minipage}{0.48\linewidth}
\begin{center}
\includegraphics[trim=0mm 0mm 0mm 0mm, clip, width=70mm]{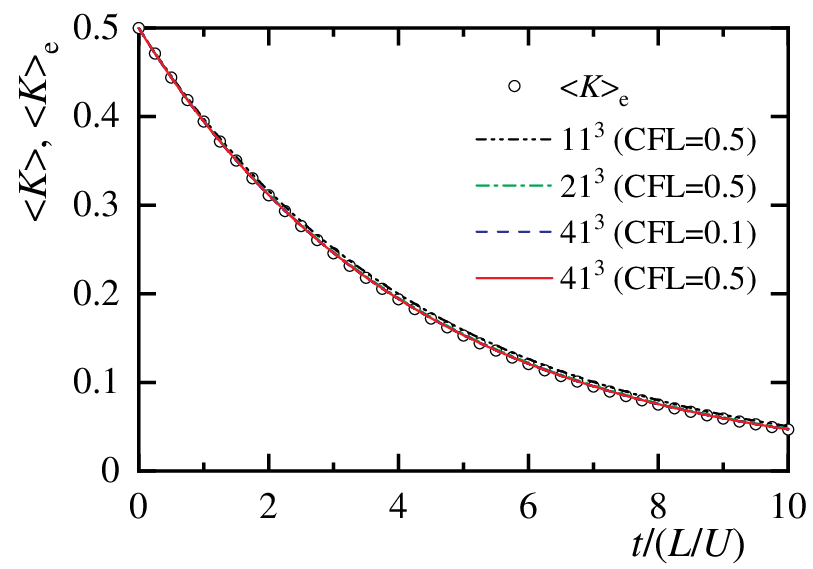} \\
(a) without initial disturbance
\end{center}
\end{minipage}
\begin{minipage}{0.48\linewidth}
\begin{center}
\includegraphics[trim=0mm 0mm 0mm 0mm, clip, width=70mm]{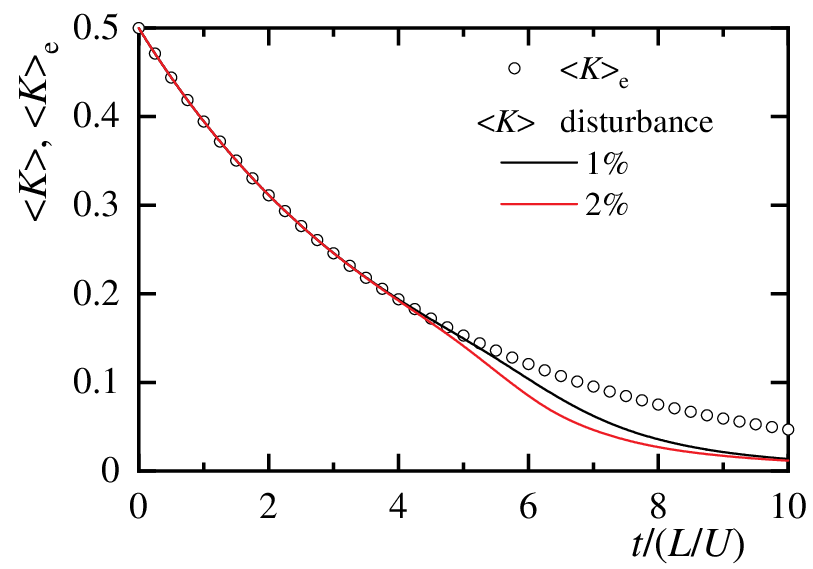} \\
(b) with initial disturbance
\end{center}
\end{minipage}
\caption{Total amounts of kinetic energy at $Re=10^3$}
\label{sum_quantity_Re1000}
\end{figure}

For $Re=10^2$, the relative error $| \varepsilon_K |$ of kinetic energy 
at the time $t/(L/U)=0.3$ after the average kinetic energy is half 
is shown in figure \ref{error_K_Re100}. 
The error is defined as $\varepsilon_K=(\langle K \rangle - \langle K \rangle_e)/\langle K \rangle_e$. 
This error also corresponds to the relative error of pressure. 
As the number of grid points increases, the error decreases with the slope of $-2$, 
indicating that the calculation is the second-order accuracy.

\begin{figure}[!t]
\begin{minipage}{0.48\linewidth}
\begin{center}
\includegraphics[trim=0mm 0mm 0mm 0mm, clip, width=70mm]{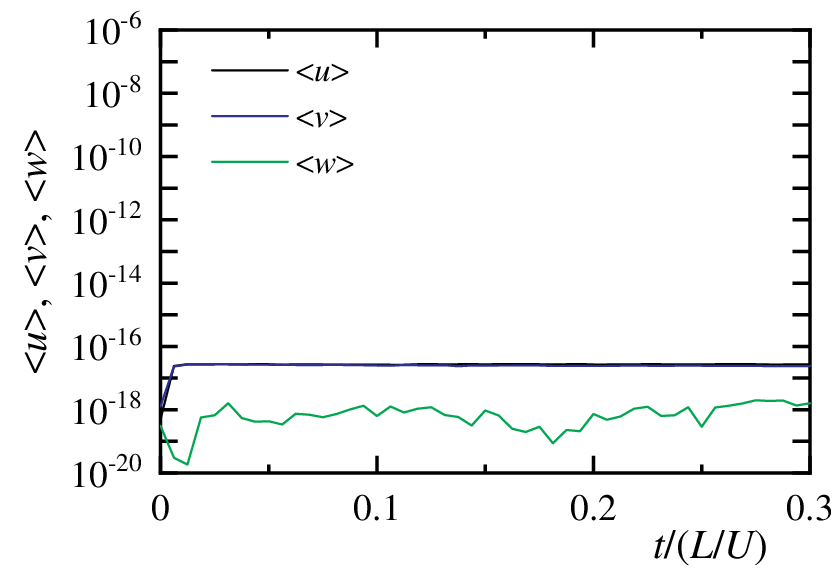} \\
(a) velocity
\end{center}
\end{minipage}
\hspace{0.02\linewidth}
\begin{minipage}{0.48\linewidth}
\begin{center}
\includegraphics[trim=0mm 0mm 0mm 0mm, clip, width=70mm]{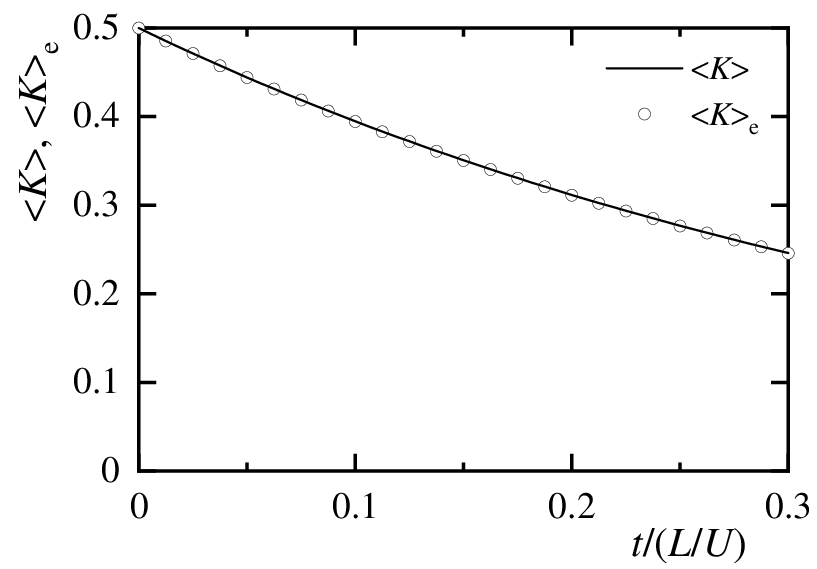} \\
(b) kinetic energy
\end{center}
\end{minipage}
%
\caption{Total amounts of velocities and kinetic energy 
under applied magnetic field: $Re=10^2$, $Re_m=1$.}
\label{sum_quantity_Re100_Rem1}
\end{figure}

Figure \ref{error_K_Re} (a) shows the time variation of the total amount 
of kinetic energy at each Reynolds number. 
Regardless of $Re$, the time variation of kinetic energy well agrees with 
the exact solution. 
Also, as $Re$ increases, the attenuation of kinetic energy slows down. 
\citet{Ham_et_al_2002} performed a calculation in which the time progress 
was inverted in an inviscid periodic flow field 
and showed the reversibility of the calculation method. 
This study set a negative time step at $Re=10^3$ 
and used the result of $t/(L/U)=3$ as an initial value. 
Then, the time marching was performed in the opposite direction. 
The result is plotted in the figure. 
Intriguingly, we can calculate the phenomenon that the decaying vortex returns to 
its original state even in the viscous fluid. 
It was possible to calculate until the time near $t/(L/U)=1.5$. 
However, as time passed, the calculation diverged on the way, 
and the decaying vortex could not be completely restored to its original state. 
In the case of Courant number 0.25, 
it was possible to calculate up to near $t/(L/U)=1$, 
but the calculation diverged on the way. 
Although this phenomenon is physically impossible, 
this time-reversal simulation may be an effective tool as a method 
for verifying the stability and conservation characteristics of a calculation method. 
Figure \ref{error_K_Re} (b) shows the relative error 
when the Reynolds number varies. 
The error maintains constant regardless of $Re$. 
Since we can confirm the calculation accuracy by changing the number of grid points 
and the Reynolds number, 
it is considered that the three-dimensional Taylor decaying vortex model 
is valuable as a benchmark test.

Next, for $Re=10^3$, the time variation of the kinetic energy 
up to the time $t/(L/U)=10$ is shown in figure \ref{sum_quantity_Re1000} (a) 
using three grids. 
In the analysis using the number of grid points of $41^3$, 
we varied the Courant number and investigated the influence of time increment. 
At $t/(L/U)=10$, the kinetic energy sufficiently attenuates. 
In the case of the number of grid points $11^3$, 
there is a slight difference between the calculation result and the analytic solution. 
On the other hand, the results obtained using finer grids well agree with the exact solution. 
Even if we perform the calculation over a long time, 
this calculated value well agrees with the exact solution, 
and there is no difference with time increments. 
In the previous study \citep{Antuono_2020}, 
the difference between the calculation result and the analytic solution increased with time, 
and the transition to turbulent flow was observed. 
This study could not confirm such a transition phenomenon at $Re=10^3$. 
Figure \ref{sum_quantity_Re1000} (b) shows the calculation result 
when an initial disturbance is added to the initial condition of the velocity. 
From around time $t/(L/U)=5$, 
the difference between the calculation result and the exact solution occurs, 
and the same trend as the existing research \citep{Antuono_2020} appears. 
We can see that increasing the initial disturbance accelerates the transition. 
For investigating the transition phenomenon of a decaying vortex, 
the analysis of a decaying vortex with an initial disturbance 
is an intriguing investigation.

\begin{figure}[!t]
\begin{minipage}{0.48\linewidth}
\begin{center}
\includegraphics[trim=0mm 0mm 0mm 0mm, clip, width=70mm]{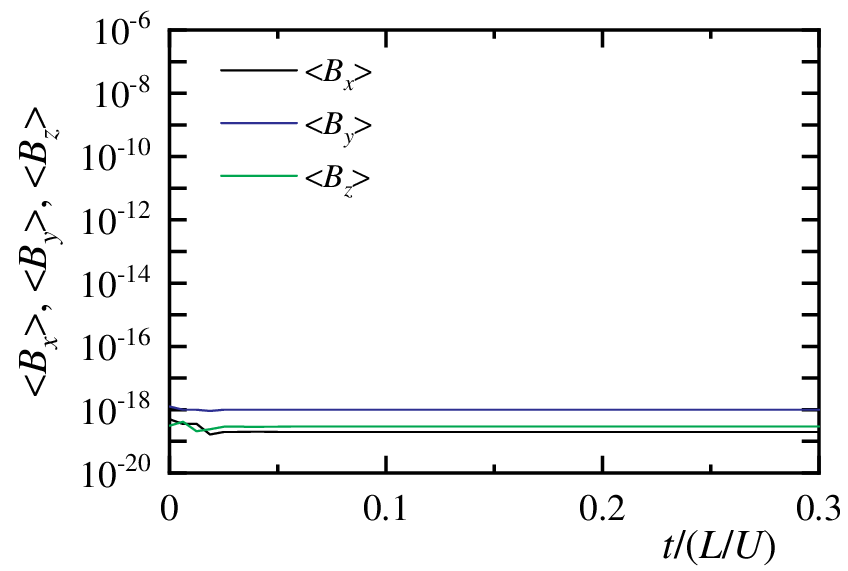} \\
(a) magnetic flux density
\end{center}
\end{minipage}
\hspace{0.02\linewidth}
\begin{minipage}{0.48\linewidth}
\begin{center}
\includegraphics[trim=0mm 0mm 0mm 0mm, clip, width=70mm]{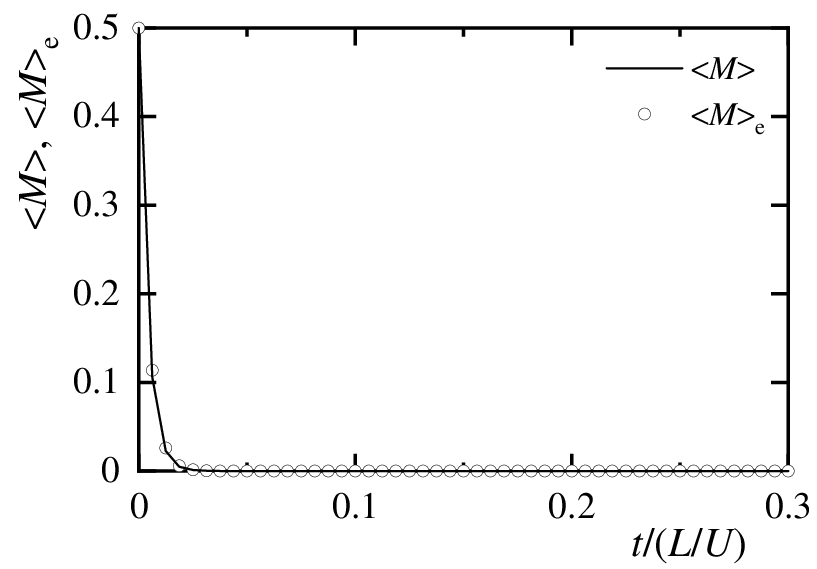} \\
(b) magnetic energy
\end{center}
\end{minipage}
\begin{minipage}{0.48\linewidth}
\begin{center}
\includegraphics[trim=0mm 0mm 0mm 0mm, clip, width=70mm]{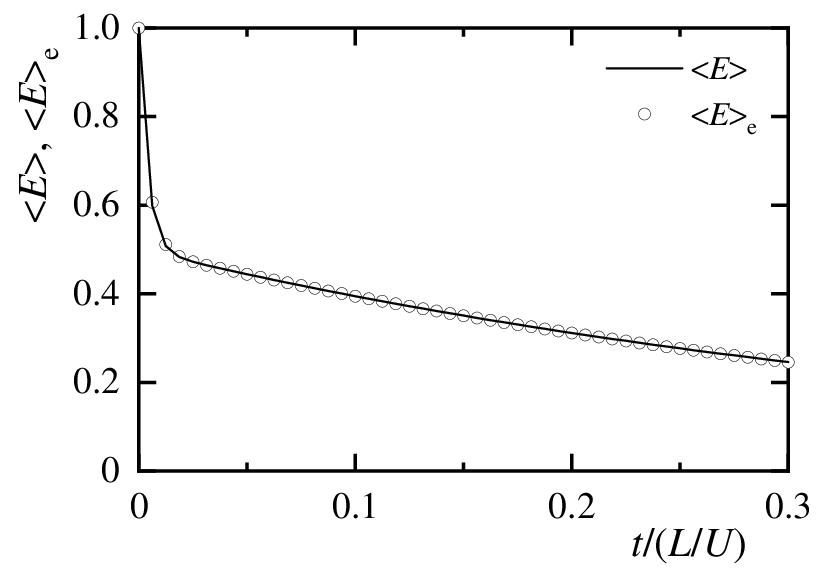} \\
(c) total energy
\end{center}
\end{minipage}
\hspace{0.02\linewidth}
\begin{minipage}{0.48\linewidth}
\begin{center}
\includegraphics[trim=0mm 0mm 0mm 0mm, clip, width=70mm]{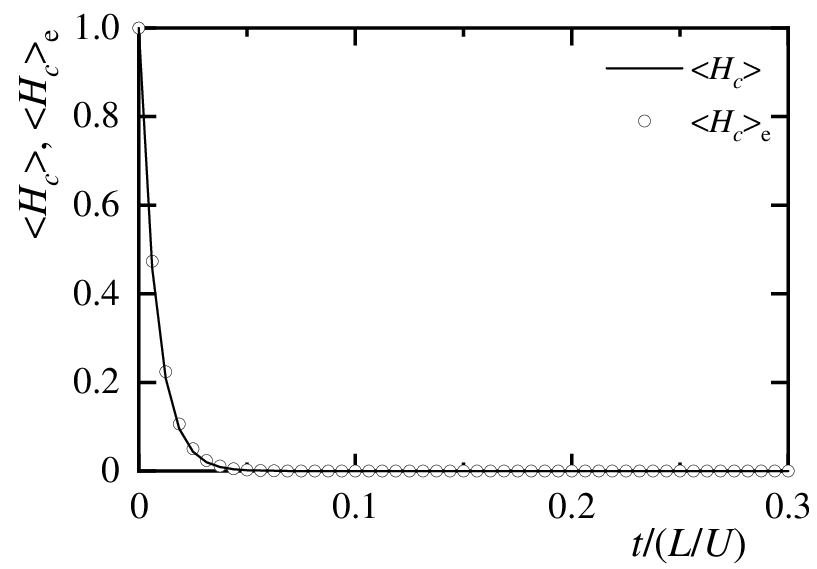} \\
(d) cross helicity
\end{center}
\end{minipage}
%
\caption{Total amounts of magnetic flux densities, magnetic energy, 
totale energy, and cross helicity 
under applied magnetic field: $Re=10^2$, $Re_m=1$.}
\label{sum_mag_quantity_Re100_Rem1}
\end{figure}

\begin{figure}[!t]
\begin{center}
\includegraphics[trim=0mm 0mm 0mm 0mm, clip, width=70mm]{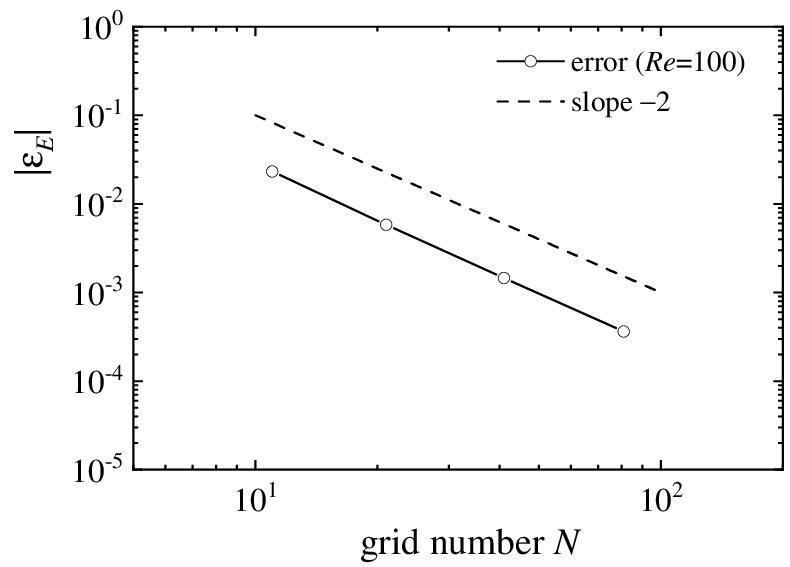} \\
\end{center}
\vspace*{-1.0\baselineskip}
\caption{Error of total energy: $Re=10^2$, $Re_m=1$, $t/(L/U)=0.3$.}
\label{error_E_Re100_Rem1}
\end{figure}

\begin{figure}[!t]
\begin{minipage}{0.48\linewidth}
\begin{center}
\includegraphics[trim=0mm 0mm 0mm 0mm, clip, width=70mm]{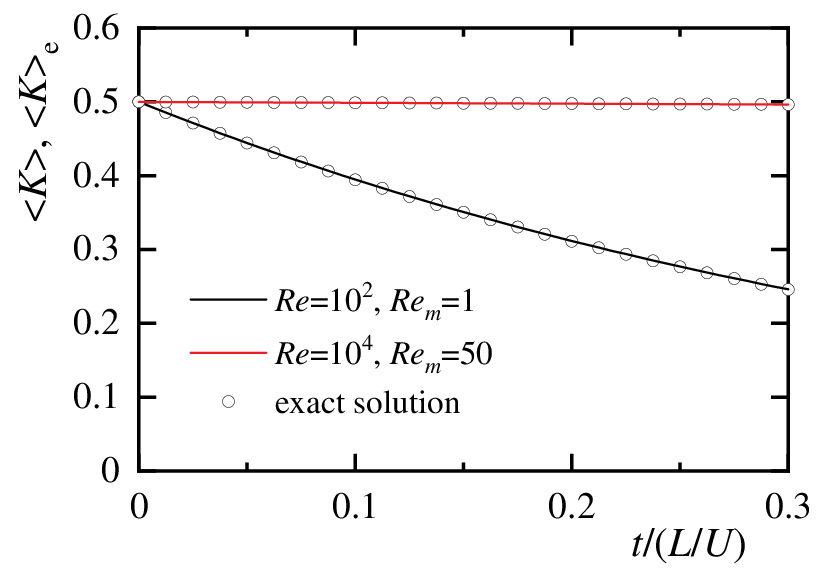} \\
(a) kinetic energy
\end{center}
\end{minipage}
\hspace{0.02\linewidth}
\begin{minipage}{0.48\linewidth}
\begin{center}
\includegraphics[trim=0mm 0mm 0mm 0mm, clip, width=70mm]{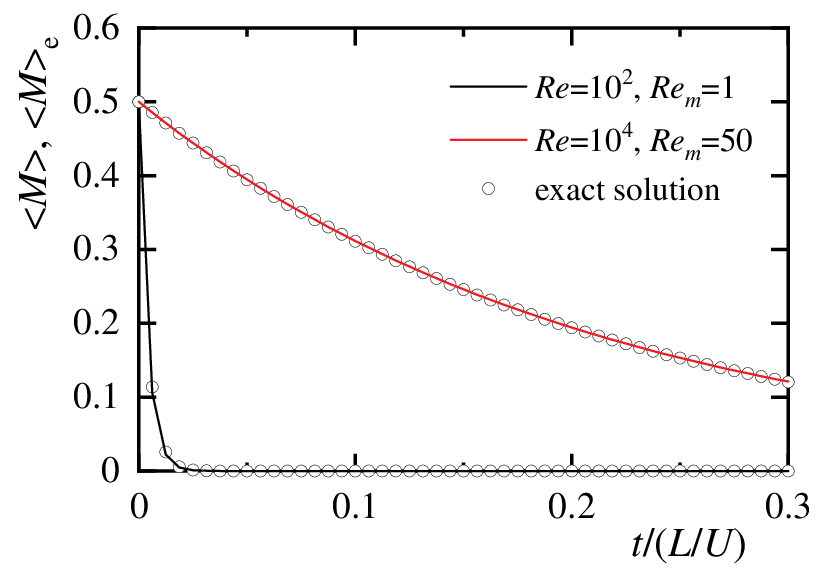} \\
(b) magnetic energy
\end{center}
\end{minipage}
\begin{minipage}{0.48\linewidth}
\begin{center}
\includegraphics[trim=0mm 0mm 0mm 0mm, clip, width=70mm]{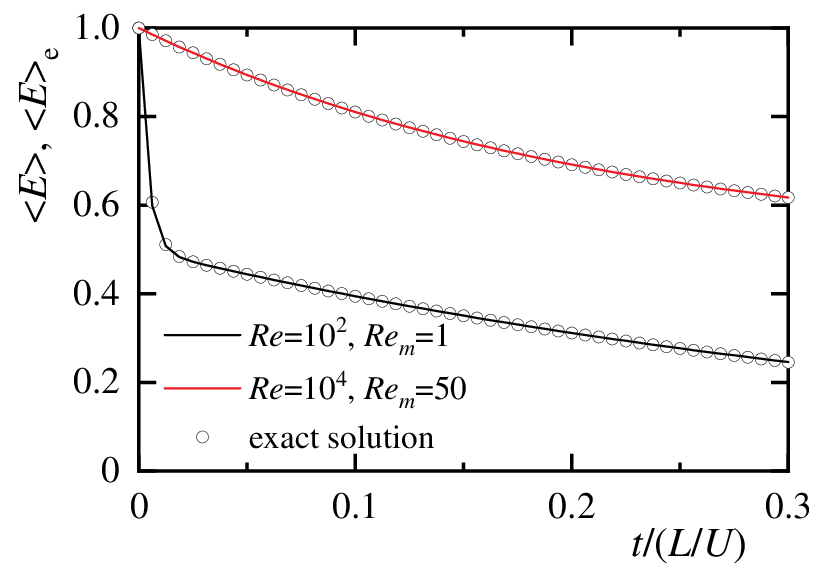} \\
(c) total energy
\end{center}
\end{minipage}
\hspace{0.02\linewidth}
\begin{minipage}{0.48\linewidth}
\begin{center}
\includegraphics[trim=0mm 0mm 0mm 0mm, clip, width=70mm]{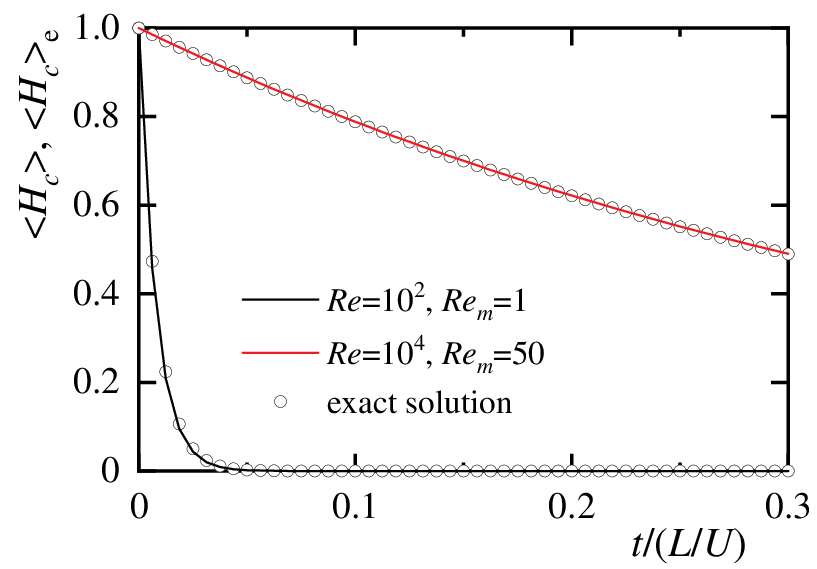} \\
(d) cross helicity
\end{center}
\end{minipage}
%
\caption{Time variations of kinetic energy, magnetic energy, 
total energy, and cross helicity 
under applied magnetic field: $Re=10^2$, $Re_m=1$, and $Re=10^4$, $Re_m=50$.}
\label{error_mag_energy}
\end{figure}

\subsection{Analysis of three-dimensional Taylor decaying vortex 
under applied magnetic field}

Figure \ref{sum_quantity_Re100_Rem1} shows the time variations of the total amounts 
of velocities and kinetic energy at $Re=10^2$. 
In addition, figure \ref{sum_mag_quantity_Re100_Rem1} shows the time variations 
of the total amounts of magnetic flux density, magnetic energy, total energy, 
and cross helicity at $Re=10^2$. 
Since the magnetic flux density is periodic, 
the total amount is conserved and analytically zero. 
The total amount of calculated magnetic flux density is the level of rounding error. 
The kinetic and magnetic energies decay with time, 
and the magnetic energy decays sharply. 
Since no Lorentz force occurs, 
we can confirm from the comparison of figures \ref{sum_quantity_Re100} 
and \ref{sum_quantity_Re100_Rem1} that the kinetic energy is consistent with 
the result with no applied magnetic field. 
If a non-physical Lorentz force occurs, 
the attenuation of kinetic energy is overestimated. 
The total energy and cross helicity agree well with each exact solution. 
Since the magnetic Reynolds number is low, the magnetic energy decays quickly, 
and most of the total energy at $t/(L/U)=0.3$ corresponds to the kinetic energy.

For $Re=10^2$, the relative error $| \varepsilon_E |$ of the total energy 
at the time $t/(L/U)=0.3$ after the average kinetic energy is half 
is shown in figure \ref{error_E_Re100_Rem1}. 
The error decreases with a slope of $-2$, 
indicating that the calculation method is the second-order accuracy.

\begin{figure}[!t]
\begin{minipage}{0.48\linewidth}
\begin{center}
\includegraphics[trim=0mm 0mm 0mm 0mm, clip, width=70mm]{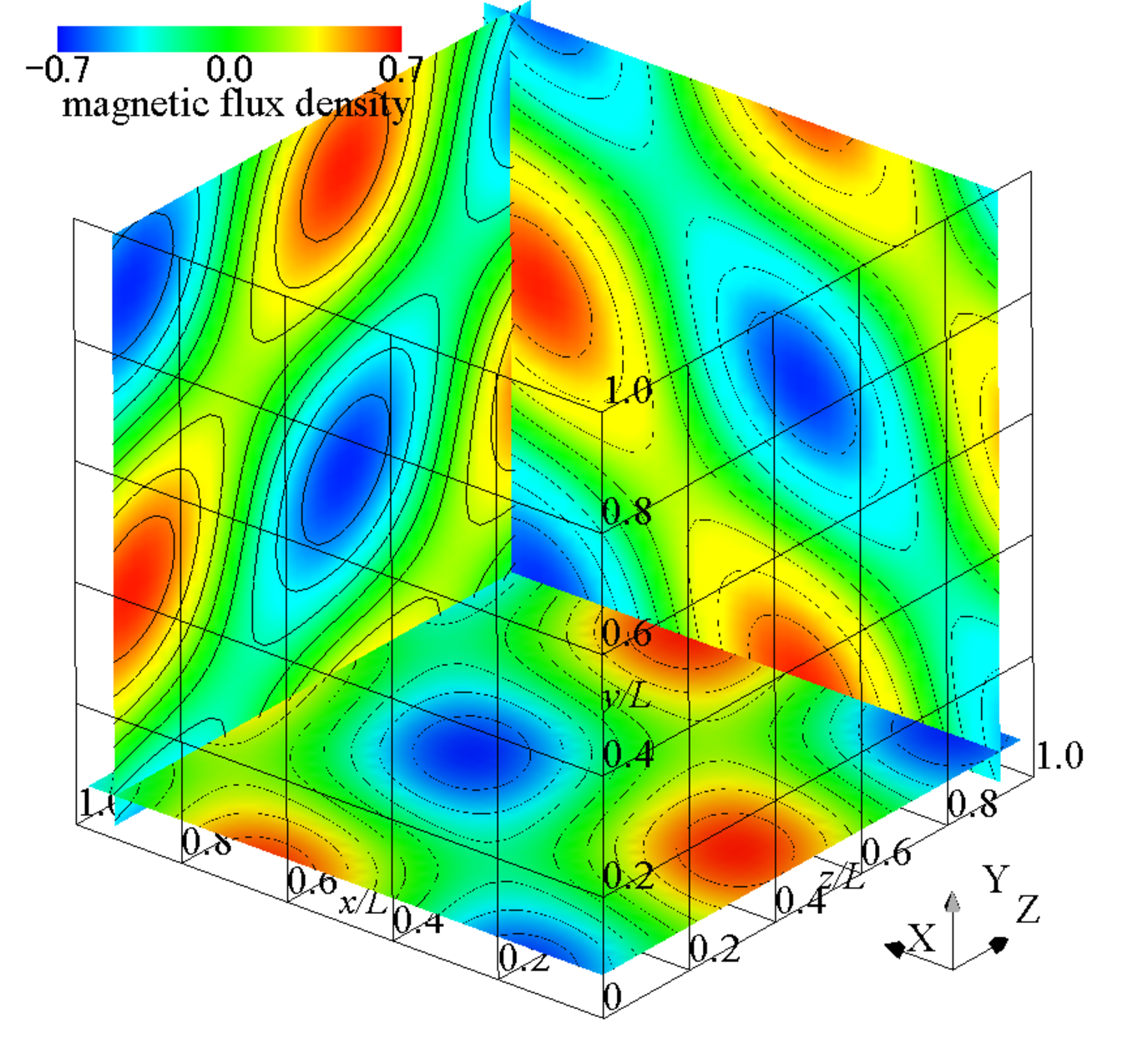} \\
(a) magnetic flux densities in $x$-, $y$-, and $z$-directions
\end{center}
\end{minipage}
\hspace{0.02\linewidth}
\begin{minipage}{0.48\linewidth}
\begin{center}
\includegraphics[trim=0mm 0mm 0mm 0mm, clip, width=70mm]{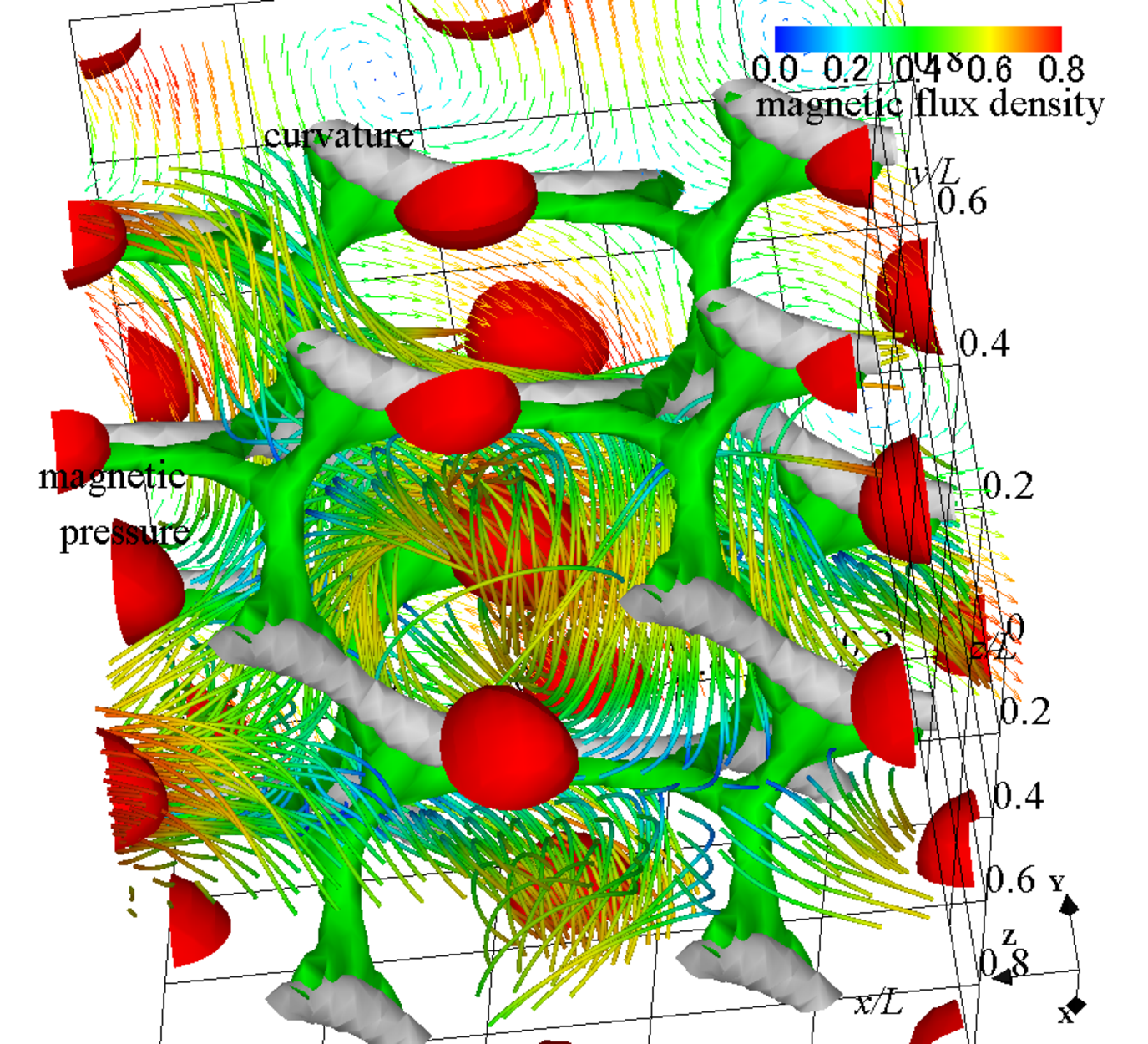} \\
(b) magnetic flux density lines, 
and isosurfaces of magnetic pressure, 2nd invariant of velocity gradient tensor, 
and curvature of magnetic pressure isosurface \\
\end{center}
\end{minipage}
\begin{center}
\includegraphics[trim=0mm 0mm 0mm 0mm, clip, width=70mm]{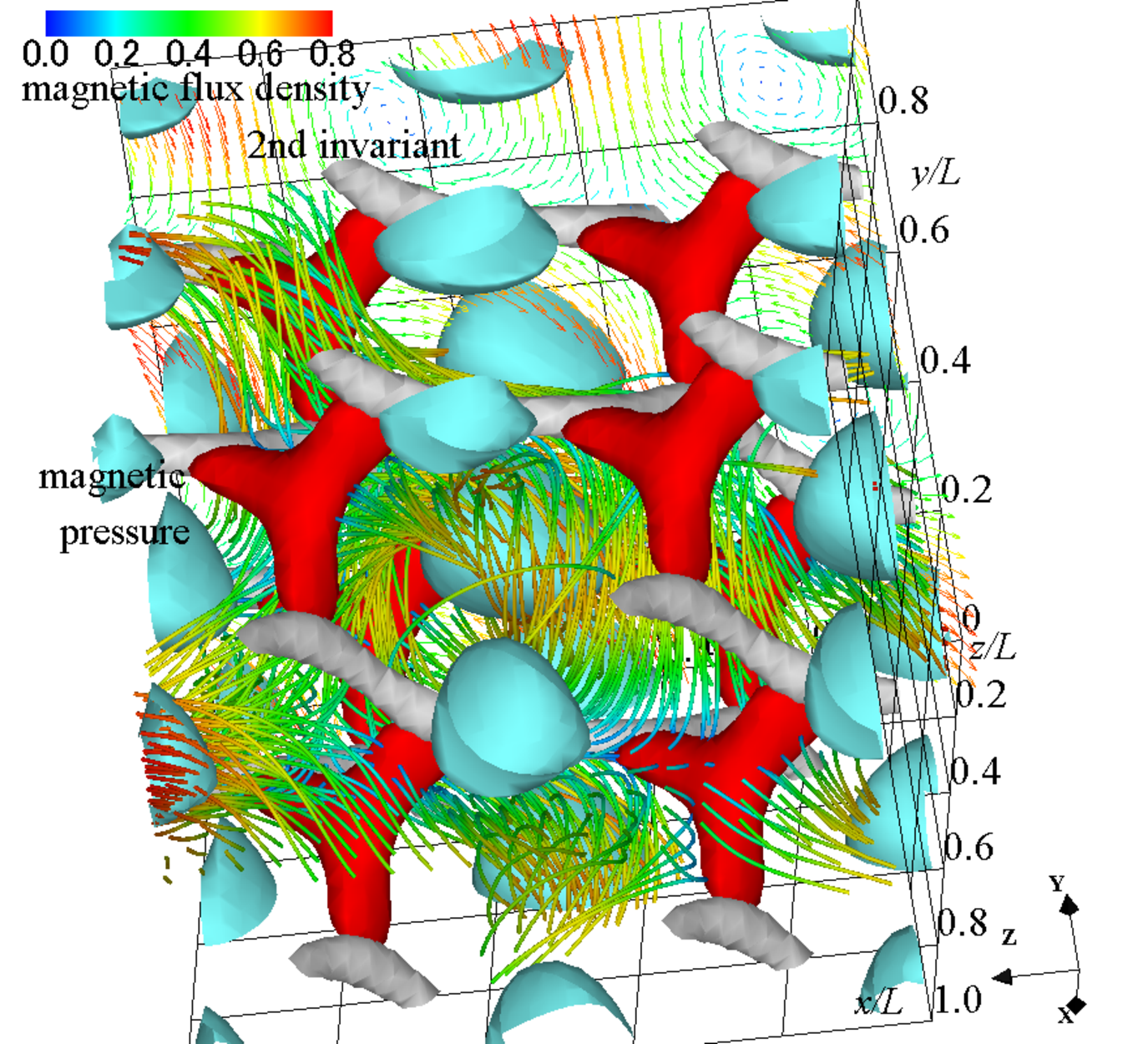} \\
(c) magnetic flux density lines, 
and isosurfaces of magnetic pressure, current density magnitude, 
and 2nd invariant of velocity gradient tensor \\
\end{center}
%
\caption{Contours of magnetic flux densities, magnetic flux density lines, 
and isosurfaces of magnetic pressure, current density magnitude, 
2nd invariant of velocity gradient tensor, and curvature of magnetic pressure isosurface: 
$Re=10^4$, $Re_m=50$, $t/(L/U)=0.3$; 
The $y$-$z$, $z$-$x$, and $x$-$y$ planes show the contours of 
magnetic flux densities in the $x$-, $y$-, and $z$-directions, respectively. 
The red, blue, silver and green isosurfaces show magnetic pressure, 
current density magnitude, 2nd invariant of velocity gradient tensor, 
and curvature of magnetic pressure isosurface, respectively}
\label{mag_flx_pres_Re10000_t03}
\end{figure}

\begin{figure}[!t]
\begin{minipage}{0.48\linewidth}
\begin{center}
\includegraphics[trim=0mm 0mm 0mm 0mm, clip, width=70mm]{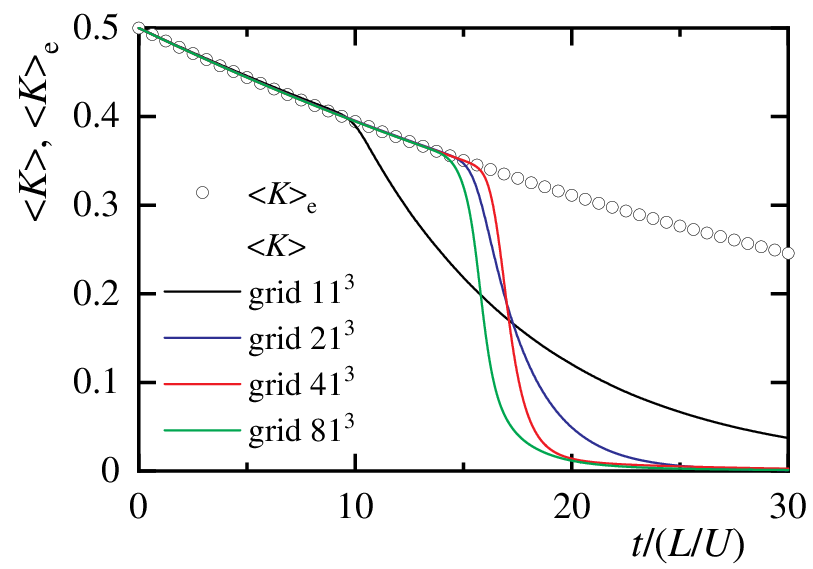} \\
(a) kinetic energy
\end{center}
\end{minipage}
\hspace{0.02\linewidth}
\begin{minipage}{0.48\linewidth}
\begin{center}
\includegraphics[trim=0mm 0mm 0mm 0mm, clip, width=70mm]{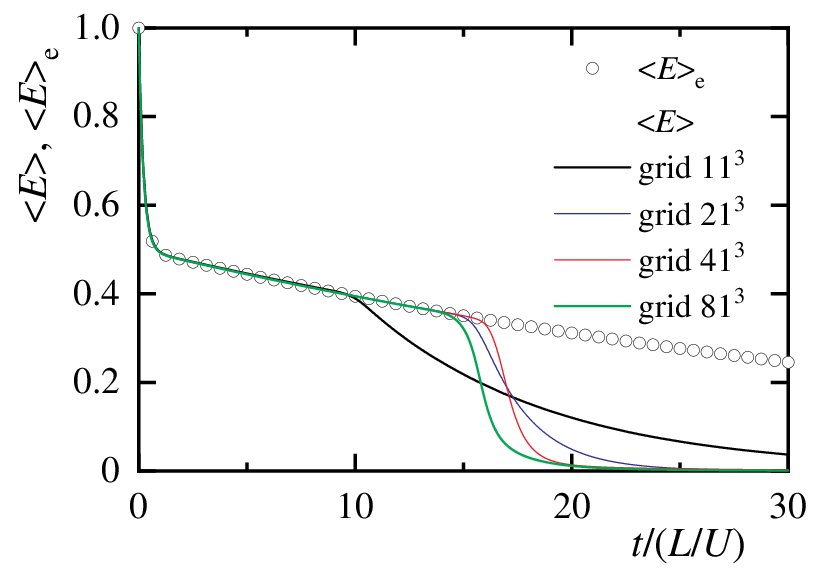} \\
(b) total energy
\end{center}
\end{minipage}
%
\caption{Time variations of kinetic energy and totale energy 
under applied magnetic field: $Re=10^4$, $Re_m=50$.}
\label{error_mag_Re10000}
\end{figure}

\begin{figure}[!t]
\begin{minipage}{0.48\linewidth}
\begin{center}
\includegraphics[trim=0mm 0mm 0mm 0mm, clip, width=70mm]{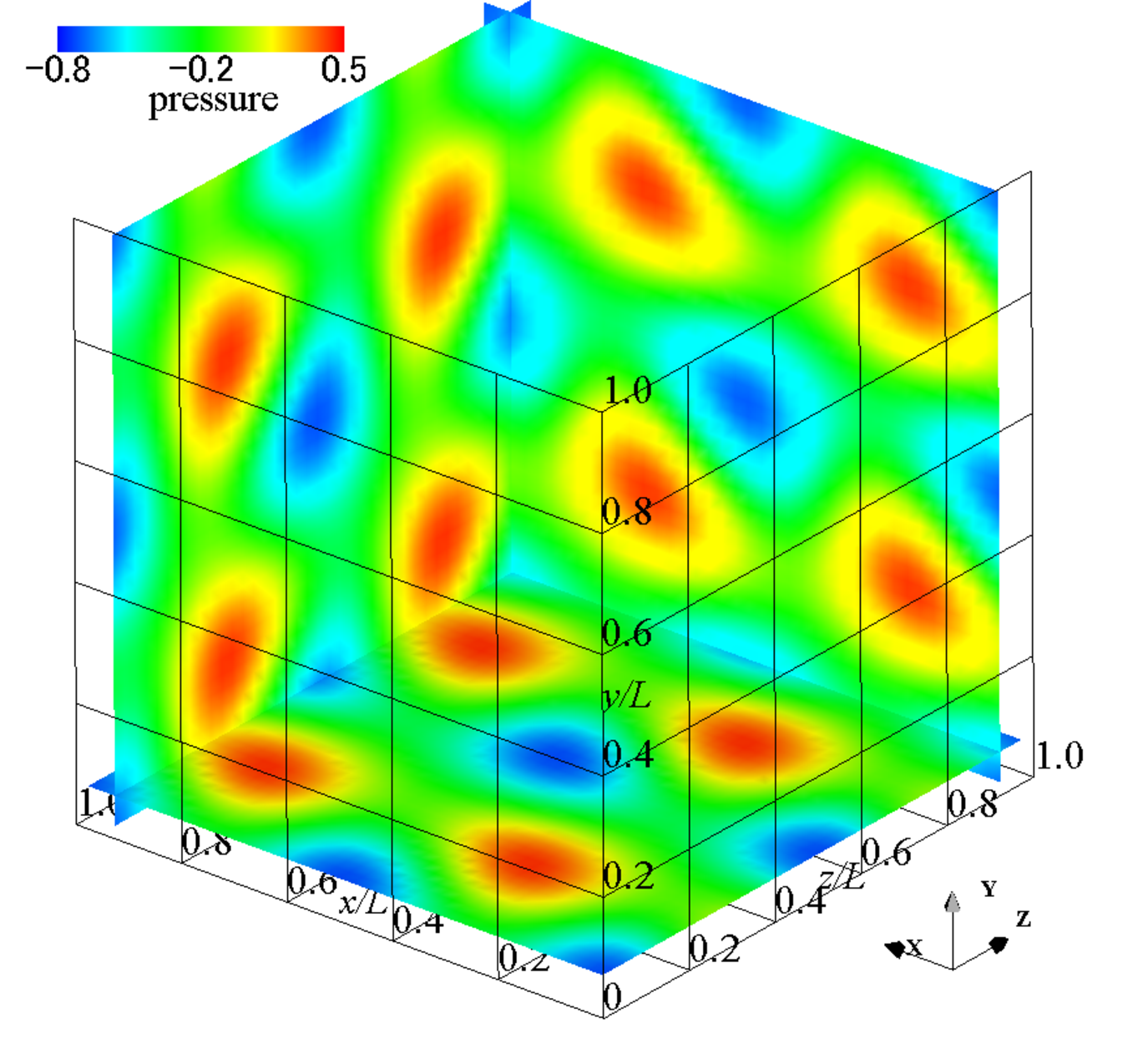} \\
(a) initial pressure
\end{center}
\end{minipage}
\hspace{0.02\linewidth}
\begin{minipage}{0.48\linewidth}
\begin{center}
\includegraphics[trim=0mm 0mm 0mm 0mm, clip, width=70mm]{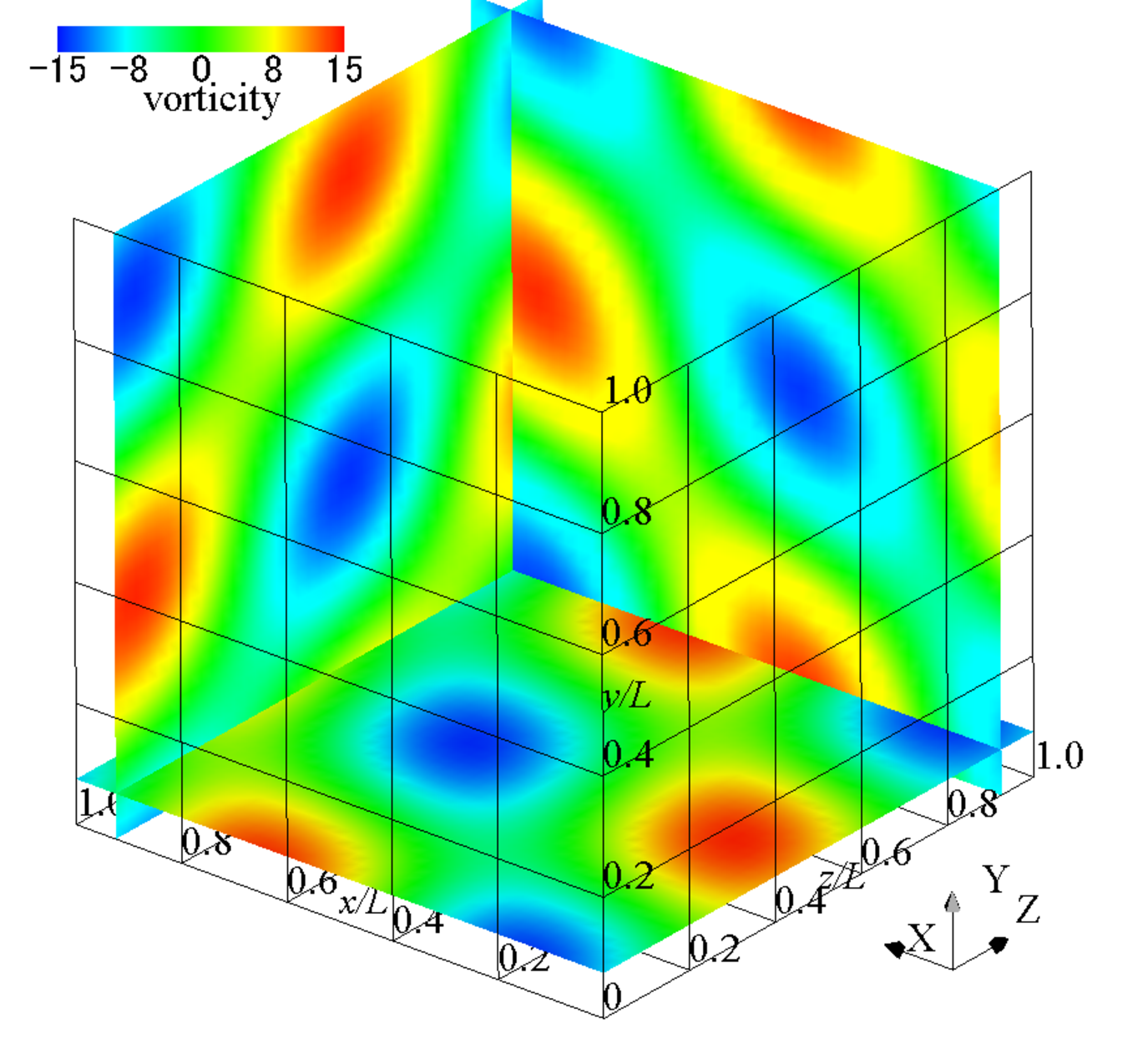} \\
(b) initial vorticity
\end{center}
\end{minipage}
\begin{minipage}{0.48\linewidth}
\begin{center}
\includegraphics[trim=0mm 0mm 0mm 0mm, clip, width=70mm]{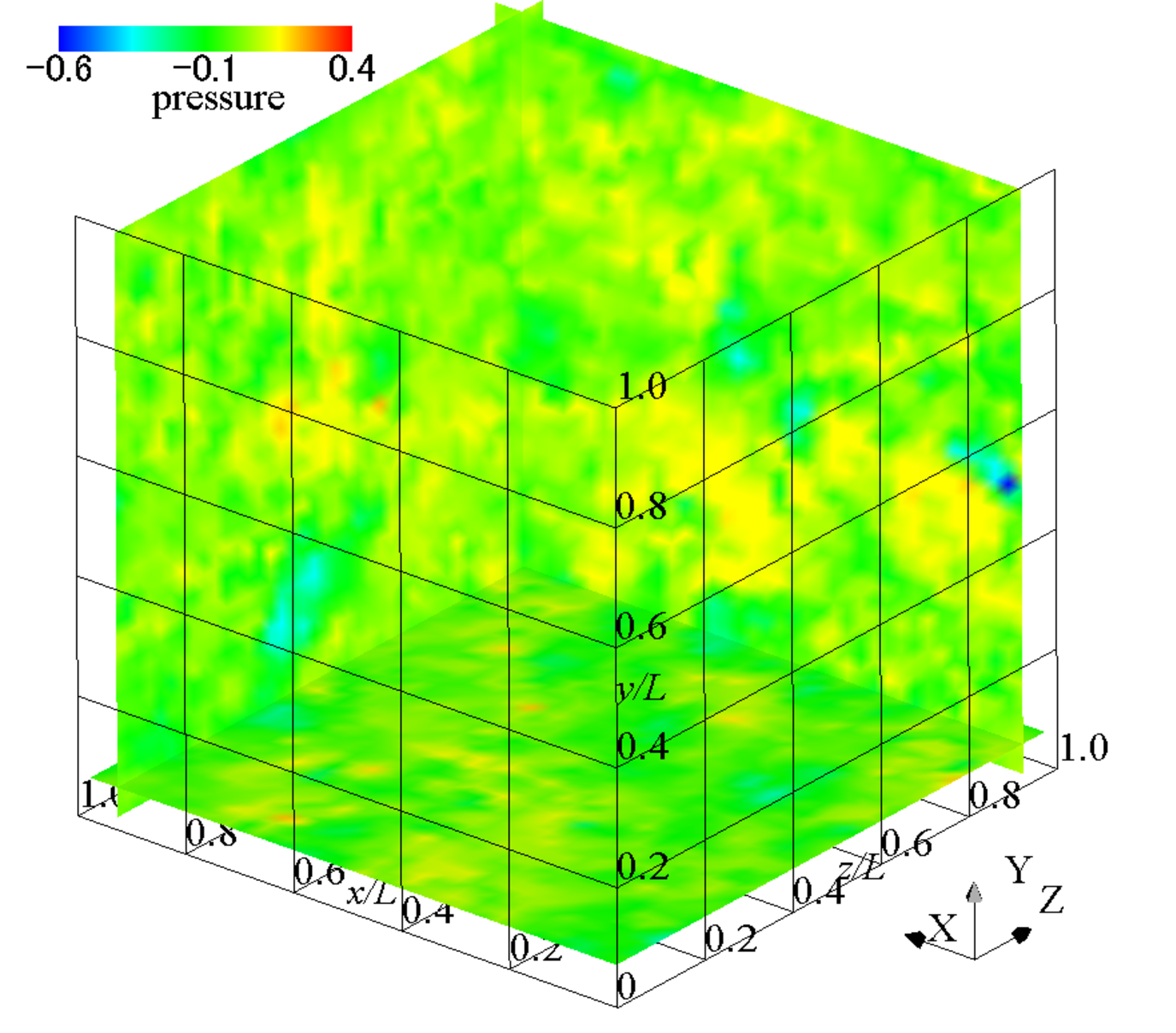} \\
(c) pressure
\end{center}
\end{minipage}
\hspace{0.02\linewidth}
\begin{minipage}{0.48\linewidth}
\begin{center}
\includegraphics[trim=0mm 0mm 0mm 0mm, clip, width=70mm]{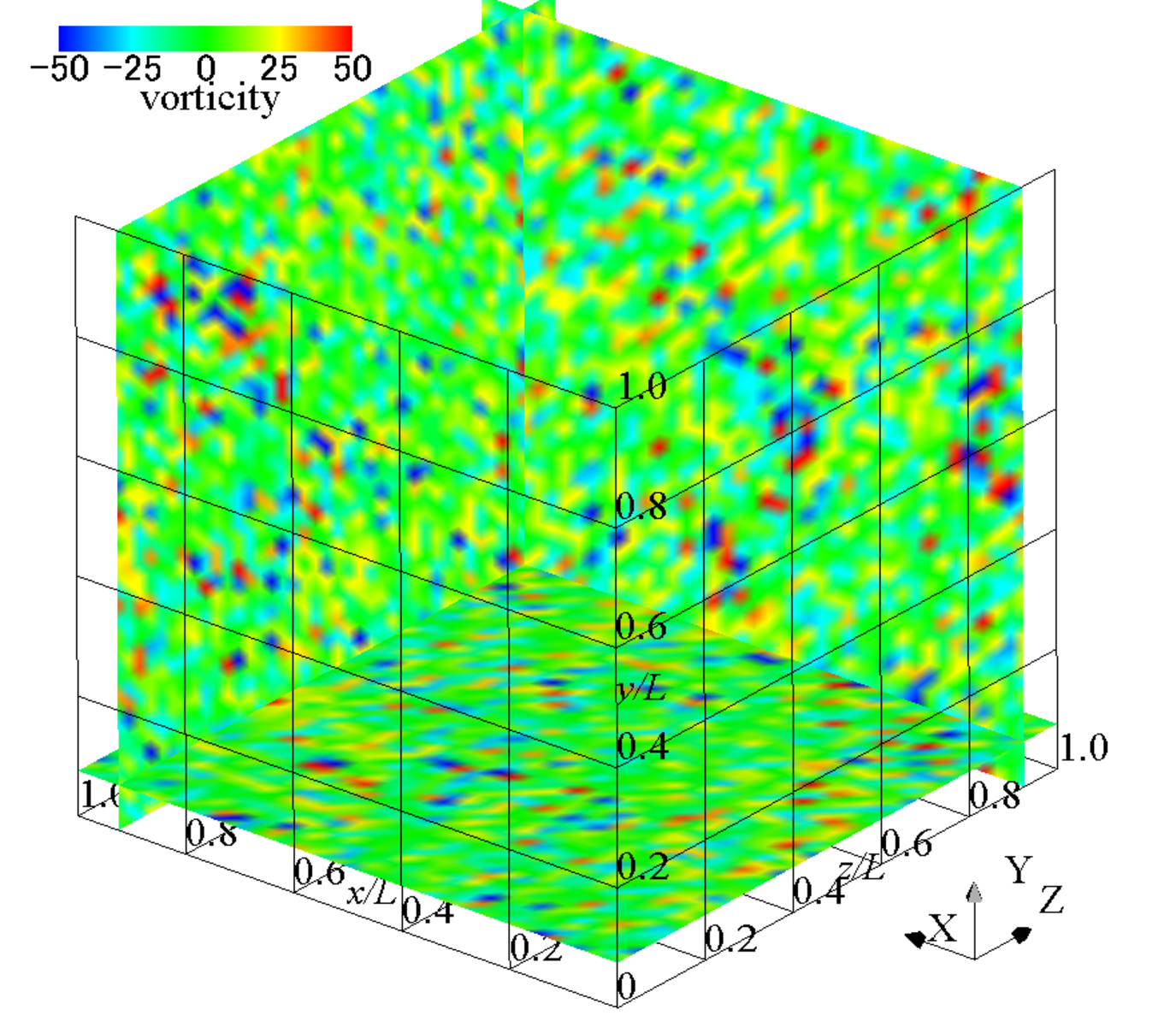} \\
(d) vorticity
\end{center}
\end{minipage}
%
\caption{Contours of pressure and vorticities under applied magnetic field: 
$Re=10^4$, $Re_m=50$, $t/(L/U)=0$, 16.85: 
The $y$-$z$, $z$-$x$, and $x$-$y$ planes show the contours of 
vorticities in the $x$-, $y$-, and $z$-directions, respectively.}
\label{pre_vor_mag_Re10000}
\end{figure}

\begin{figure}[!t]
\begin{center}
\includegraphics[trim=0mm 0mm 0mm 0mm, clip, width=70mm]{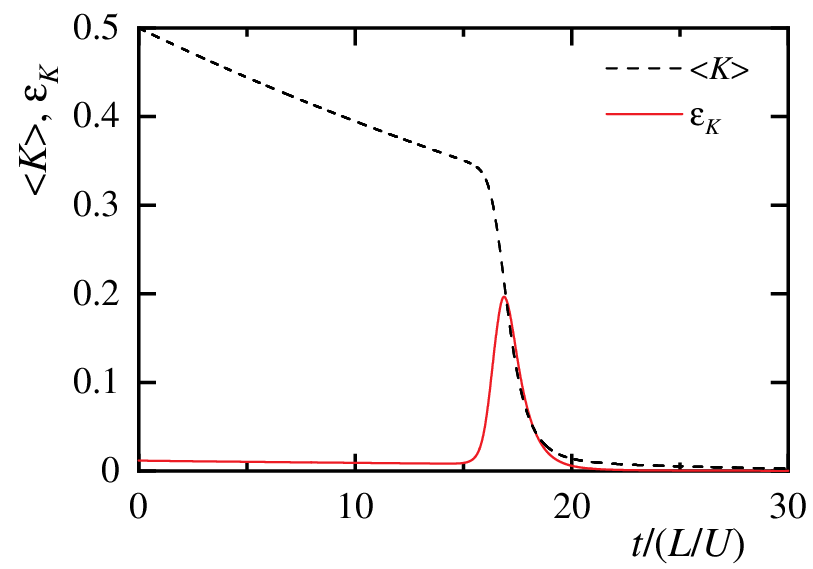} \\
\end{center}
\vspace*{-1.0\baselineskip}
\caption{Time variations of kinetic energy and energy dissipation
under applied magnetic field: $Re=10^4$, $Re_m=50$.}
\label{energy_dissipation}
\end{figure}

Next, we investigate the trend of the decaying vortex 
when the calculation conditions vary. 
Figure \ref{error_mag_energy} shows the time variations in the kinetic, magnetic, 
and total energies and cross helicity at $Re=10^4$ and $Re_m=50$. 
For comparison, the results of $Re=10^2$ and $Re_m=1$ are also included. 
This calculation result agrees well with the exact solution, 
and the present analysis can capture the energy decay process. 
Under $Re=10^4$ and $Re_m=50$, there is almost no attenuation of the kinetic energy. 
On the other hand, the magnetic energy, total energy, and cross helicity decay over time.

Figure \ref{mag_flx_pres_Re10000_t03} (a) shows the distribution 
of the magnetic flux density at time $t/(L/U)=0.3$. 
The $y$-$z$, $z$-$x$, and $x$-$y$ cross-sections show the distribution 
of magnetic flux densities in the $x$-, $y$-, and $z$-directions, respectively. 
At this time, the distinct periodicity of the magnetic flux density remains. 
In addition, figures \ref{mag_flx_pres_Re10000_t03} (b) and (c) show 
the magnetic flux lines, the magnetic pressure, 
the magnitude of the current density vector, the second invariant of the velocity gradient tensor, 
and the curvature of the isosurface of magnetic pressure. 
Here, we calculated the curvature of the isosurface of magnetic pressure 
to visualise the region of low magnetic pressure. 
The red isosurfaces show the distributions of magnetic pressure $P=0.01$ and $P=0.28$, 
and the light blue isosurface expresses the current density magnitude $J=7.5$, 
indicating the occurrence of high current density. 
The silver isosurface shows the second invariant $Q=-53$ of the velocity gradient tensor 
and represents a tubular high-shear region. 
The green isosurface shows the curvature $\kappa=-50$ of the magnetic pressure isosurface, 
and we can confirm the region of low magnetic pressure. 
The high magnetic pressure is shown in figure (b), 
and the low magnetic pressure and the magnitude of high current density 
are shown in figure (c). 
In figure (b), the magnetic pressure distribution of a distorted cubic structure 
appears so that the regions of low magnetic pressure are connected 
in a mesh pattern and surround the areas of high magnetic pressure. 
At this time, the attenuation of the velocity field is small, 
so the clear vortex structure is present. 
The magnetic flux density decays, but the magnetic flux lines are similar to 
the streamlines of the velocity field. 
The distribution of magnetic pressure shown in figure (c) is Y-shaped 
and the same as the shape of the pressure distribution in figure \ref{flow_Re100}. 
The dimensionless magnetic pressure corresponds to the dimensionless magnetic energy. 
Therefore, the magnetic energy becomes high in the region, 
where the magnetic pressure is high.
The high current density occurs in a grid pattern, 
and the magnetic flux lines swirl so as to surround the high current density region. 
In the area of high current density, the magnetic pressure, 
that is, the magnetic energy becomes high.

Figure \ref{error_mag_Re10000} shows the time variations in the kinetic 
and total energies at $Re=10^4$ and $Re_m=50$. 
Over time, the difference between the result obtained using each grid and 
the exact solution appears, 
and the kinetic and total energies decay sharply. 
An existing study \citep{Antuono_2020} reported that the difference 
from the analytic solution suggested a transition to turbulent flow. 
When the flow transition occurred, the magnetic energy had sufficiently decreased. 
Therefore, under this condition, 
the influence of the magnetic field on the flow transition is considered small. 
As the number of grid points increases, 
the transition points approach the dimensionless time $t/(L/U)=16$.

The vorticity distribution at time $t/(L/U)=0$ and 16.85 is shown 
in figure \ref{pre_vor_mag_Re10000}. 
The $y$-$z$, $z$-$x$, and $x$-$y$ cross-sections show the vorticity distributions 
in the $x$-, $y$-, and $z$-directions, respectively. 
In the initial state, a large-scale vortex exists, 
but at $t/(L/U)=16.85$, it can be seen from the figure 
that it is converted to small-scale vortex structures by a non-linear effect 
and attenuated.

Figure \ref{energy_dissipation} shows the time variation of the kinetic energy $K_{av}$ 
and its dissipation rate $\varepsilon_K$. 
The dissipation rate is defined as $\varepsilon_K=-d K_{av}/dt$. 
As the kinetic energy decays sharply, 
the dissipation rate increases and a maximum value appears. 
The pressure and vorticity distributions 
when the dissipation rate is maximum were shown 
in figures \ref{pre_vor_mag_Re10000} (c) and (d). 
The vortex structure disappears with time due to viscous dissipation, 
the induced magnetic field also disappears, 
and the flow field asymptotically approaches the stationary state.

\section{Conclusions}

In this study, we analysed a three-dimensional Taylor decaying vortex 
under an applied magnetic field 
and investigated the effectiveness of the decaying magnetic field model 
for verifying the calculation method of electromagnetic fluid flow. 
The following findings were obtained.

First, we investigated the characteristics of a three-dimensional Taylor decaying vortex 
without an applied magnetic field. 
In the flow field where the decaying vortex exists, 
high-pressure regions are connected in a mesh pattern 
so as to include stagnation points, 
and the pressure distribution with a distorted cubic structure appears 
around a low-pressure region. 
A tubular high-shear structure, which passes through the cube structure, exists, 
and a swirling flow forms around the low-pressure region inside the cube. 
Furthermore, we investigated the total amounts of velocity, pressure, 
and kinetic energy when the number of grid points and the Reynolds number varied 
and clarified the conservation characteristics of the transport quantities 
in the three-dimensional Taylor decaying vortex.

Next, we analysed a three-dimensional Taylor decaying vortex 
under an applied magnetic field 
and clarified the characteristics of the decaying magnetic field. 
When a magnetic field is applied, 
low magnetic pressure regions are connected in a mesh pattern, 
and the magnetic pressure distribution with a distorted cubic structure occurs 
to surround a high magnetic pressure region. 
In a stagnation region, the magnetic energy becomes low, 
and the magnetic flux line is similar to the streamline of the velocity field. 
High current densities occur in a grid pattern, 
and the magnetic flux lines swirl around the high current density region. 
The magnetic pressure and magnetic energy are high in the high current density region. 
When the Reynolds number and the magnetic Reynolds number vary, 
the decay trends of various energies agree well with the exact solution. 
The transition to turbulent flow occurs at a high Reynolds number, 
and the kinetic and total energies decrease rapidly. 
After the dissipation rate of kinetic energy becomes maximum, 
the vortex structure decays, 
and the flow field gradually approaches a stationary state without magnetic fields.

The three-dimensional decaying magnetic field belonging to the Beltrami flow 
is a simple model for investigating the energy conservation characteristics 
of a calculation method 
and is also an intriguing target for studying the transition to turbulent flow 
and energy dissipation. 
This model of the three-dimensional decaying magnetic field is considered valuable 
as one benchmark test.

\appendix
\section{Discretization of the Lorentz force}
\label{subsection_Lorentz}

This study uses a staggered grid \citep{Amsden&Harlow_1970}. 
We denote the cell centre as $(i, j, k)$. 
The magnetic flux densities, $B_x$, $B_y$, and $B_z$, are defined 
at the cell interface $(i+1/2, j, k)$, $(i, j+1/2, k)$, 
and $(i, j, k+1/2)$, respectively, as with the velocity field. 
The coordinate transformation is performed so that equation can be discretized 
on non-uniform grids. 
As an example, the Lorentz force in the $x$-direction can be transformed 
as follows:
\begin{equation}
   \left. F_x \right|_{i+1/2,j,k} 
   = \frac{1}{Al^2} \frac{1}{\bar{J}^{\xi}_{i+1/2,j,k}} \left( 
     \overline{J \bar{B_z}^{\xi} J_y}^{\zeta}_{i+1/2,j,k} 
   - \overline{J \bar{B_y}^{\xi} J_z}^{\eta}_{i+1/2,j,k} \right),
   \label{Lorentz1}
\end{equation}
where $J$ is the Jacobian defined as $J = x_{\xi} y_{\eta} z_{\zeta}$. 
The Cartesian coordinate $(x, y, z)$ in the physical space is transformed 
into the computational space $(\xi, \eta, \zeta)$. 
$\bar{B_y}^{\xi}$ and $\bar{B_z}^{\xi}$ are interpolated values. 
The variable at grid point $(i, j, k)$ is defined as $\Psi_{i,j,k}$. 
The interpolation in the $x$ ($\xi$)-direction for the variable $\Psi$ 
is given as
\begin{equation}
   \left. \bar{\Psi}^{\xi} \right|_{i+1/2,j,k} 
   = \frac{\Psi_{i,j,k} + \Psi_{i+1,j,k}}{2}.
\end{equation}
In this study, the current densities, $J_y$ and $J_z$, are not defined 
at the same definition points as the velocities in the $y$- and $z$-directions, respectively. 
$J_x$, $J_y$, and $J_z$ are defined at the midpoints of the cell edges, 
$(i, j+1/2, k+1/2)$, $(i+1/2, j, k+1/2)$, and $(i+1/2, j+1/2, k)$, respectively. 
In this case, 2 surrounding magnetic flux densities $B_x$ are required 
to calculate $J_{y i+1/2,j,k+1/2}$. 
The 2 surrounding current densities $J_{y i+1/2,j,k+1/2}$, are required 
to calculate $F_{x i+1/2,j,k}$. 
A total of 3 surrounding $B_x$ is required. 
In this way, the Lorentz force can be obtained by compact interpolation. 
In addition, the charge conservation law is satisfied 
at the grid point $(i+1/2, j+1/2, k+1/2)$. 
This method is the compact interpolation \citep{Yanaoka_2022}.

The Lorentz force is rewritten using compact interpolation 
from the non-conservative to the conservative form. 
The compatibility between conservative and non-conservative forms is preserved 
in discretized equations for uniform grids 
and approximately for non-uniform grids. 
Therefore, even if the non-conservative Lorentz force is discretized, 
the momentum and kinetic energy are conserved in inviscid fluids.


\vspace*{1.0\baselineskip}
\noindent
{\bf Acknowledgements.}
This research did not receive any specific grant from funding agencies 
in the public, commercial, or not-for-profit sectors. 
We would like to express our gratitude to Associate Professor Yosuke Suenaga 
of Iwate University for his support of our laboratory. 
The authors wish to acknowledge the time and effort of everyone involved in this study.

\vspace*{1.0\baselineskip}
\noindent
{\bf Declaration of interests.}
The authors report no conflicts of interest.

\vspace*{1.0\baselineskip}
\noindent
{\bf Author ORCID.} \\
H. Yanaoka \url{https://orcid.org/0000-0002-4875-8174}.


\bibliographystyle{arXiv_elsarticle-harv}
\bibliography{decaying_vortex_bibfile}

\end{document}